\documentclass[longbibliography,aps,reprint,groupedaddress,superscriptaddress]{revtex4-2}

\usepackage{graphicx}
\usepackage{bm}
\usepackage{amsmath,amssymb}
\usepackage{hyperref}
\usepackage{lipsum}
\usepackage{color}
\usepackage{soul}
\usepackage{bm}
\usepackage{textcase}
\usepackage{textcomp}
\usepackage{braket}
\usepackage{setspace}
\usepackage{mathtools}

\definecolor{colorref}{rgb}{0.0, 0.408, 0.647}
\definecolor{grey}{rgb}{0.95, 0.95, 0.95}
\hypersetup{
	colorlinks = true,
	allcolors = colorref
}

\definecolor{colora}{rgb}{0.8, 0.0, 0.2}

\newcommand{\figtitle}[1]{\textbf{#1}}
\newcommand{\subfiglabel}[1]{\textbf{#1}}

\newcommand{\rfig}[1]{Fig.~\textcolor{colorref}{\ref{#1}}}

\newcommand{\SeeSupply}[1]{Supplemental Information}

\newcommand{\SIQSE}{\affiliation{1}{Shenzhen Institute for Quantum Science and Engineering, Southern University of Science and Technology, Shenzhen 518055, China}}

\newcommand{\IQA}{\affiliation{3}{International Quantum Academy, Shenzhen 518048, China}}
\newcommand{\GDKL}{\affiliation{4}{Guangdong Provincial Key Laboratory of Quantum Science and Engineering, Southern University of Science and Technology, Shenzhen 518055, China}}

\newcommand{\HFNL}{\affiliation{6}{
Shenzhen Branch, Hefei National Laboratory, Shenzhen 518048, China}}
\newcommand{\NXU}{\affiliation{7}{
School of Physics, Ningxia University, Yinchuan 750021, PR China}}

\begin{document}

\title{Logical multi-qubit entanglement with dual-rail superconducting qubits}
\author{Wenhui Huang}
\thanks{These authors contributed equally to this work.}
\affiliation{\IQA}\affiliation{\SIQSE}\affiliation{\GDKL}
\author{Xuandong Sun}
\thanks{These authors contributed equally to this work.}
\affiliation{\SIQSE}\affiliation{\IQA}\affiliation{\GDKL}
\author{Jiawei Zhang}
\thanks{These authors contributed equally to this work.}
\affiliation{\SIQSE}\affiliation{\IQA}\affiliation{\GDKL}

\author{Zechen Guo}
\affiliation{\SIQSE}\affiliation{\IQA}\affiliation{\GDKL}

\author{Peisheng Huang}
\affiliation{\NXU}\affiliation{\IQA}

\author{Yongqi Liang}
\affiliation{\SIQSE}\affiliation{\IQA}\affiliation{\GDKL}

\author{Yiting Liu}
\affiliation{\SIQSE}\affiliation{\IQA}\affiliation{\GDKL}

\author{Daxiong Sun}
\affiliation{\SIQSE}\affiliation{\IQA}\affiliation{\GDKL}

\author{Zilin Wang}
\affiliation{\NXU}\affiliation{\IQA}

\author{Yuzhe Xiong}
\affiliation{\SIQSE}\affiliation{\IQA}\affiliation{\GDKL}

\author{Xiaohan Yang}
\affiliation{\SIQSE}\affiliation{\IQA}\affiliation{\GDKL}

\author{Jiajian Zhang}
\affiliation{\SIQSE}\affiliation{\IQA}\affiliation{\GDKL}

\author{Libo Zhang}
\affiliation{\SIQSE}\affiliation{\IQA}\affiliation{\GDKL}

\author{Ji Chu}
\affiliation{\IQA}

\author{Weijie Guo}
\affiliation{\IQA}

\author{Ji Jiang}
\affiliation{\SIQSE}\affiliation{\IQA}\affiliation{\GDKL}

\author{Song Liu}
\affiliation{\SIQSE}\affiliation{\IQA}\affiliation{\GDKL}\affiliation{\HFNL}

\author{Jingjing Niu}
\affiliation{\IQA}\affiliation{\HFNL}

\author{Jiawei Qiu}
\affiliation{\IQA}

\author{Ziyu Tao}
\affiliation{\IQA}

\author{Yuxuan Zhou}
\affiliation{\IQA}

\author{Xiayu Linpeng}
\email{linpengxiayu@iqasz.cn}
\affiliation{\IQA}

\author{Youpeng Zhong}
\email{zhongyoupeng@iqasz.cn}
\affiliation{\IQA}\affiliation{\HFNL}

\author{Dapeng Yu}
\affiliation{\IQA}\affiliation{\HFNL}

\date{\today}

\begin{abstract}
Recent advances in quantum error correction (QEC) across hardware platforms have demonstrated operation near and beyond the fault-tolerance threshold~\cite{Chen2021,Acharya2023,Acharya2024}, yet achieving exponential suppression of logical errors through code scaling remains a critical challenge. Erasure qubits, which enable hardware-level detection of dominant error types, offer a promising path toward resource-efficient QEC by exploiting error bias~\cite{Grassl1997,Bennett1997,Cong2022,Kang2023,Kubica2023}.
Single erasure qubits with dual-rail encoding in superconducting cavities and transmons have demonstrated high coherence and low single-qubit gate errors with mid-circuit erasure detection~\cite{Levine2024,Koottandavida2024,Chou2024,Graaf2025}, but the generation of multi-qubit entanglement--a fundamental requirement for quantum computation and error correction--has remained an outstanding milestone. Here, we demonstrate a superconducting processor integrating four dual-rail erasure qubits that achieves the logical multi-qubit entanglement with error-biased protection. Each dual-rail qubit, encoded in pairs of tunable transmons, preserves millisecond-scale coherence times and single-qubit gate errors at the level of $10^{-5}$. By engineering tunable couplings between logical qubits, we generate high-fidelity entangled states resilient to physical qubit noise, including logical Bell states (98.8\% fidelity) and a three-logical-qubit Greenberger-Horne-Zeilinger (GHZ) state (93.5\% fidelity). A universal gate set is realized through a calibrated logical controlled-NOT (CNOT) gate with 96.2\% process fidelity, enabled by coupler-activated $XX$ interactions in the protected logical subspace.
This work advances dual-rail architectures beyond single-qubit demonstrations, providing a blueprint for concatenated quantum error correction~\cite{Putterman2025} with erasure qubits. 

\end{abstract}
\maketitle

\section{Introduction}

\begin{figure*}[!t]
    \centering
    \includegraphics[width=0.65\textwidth]{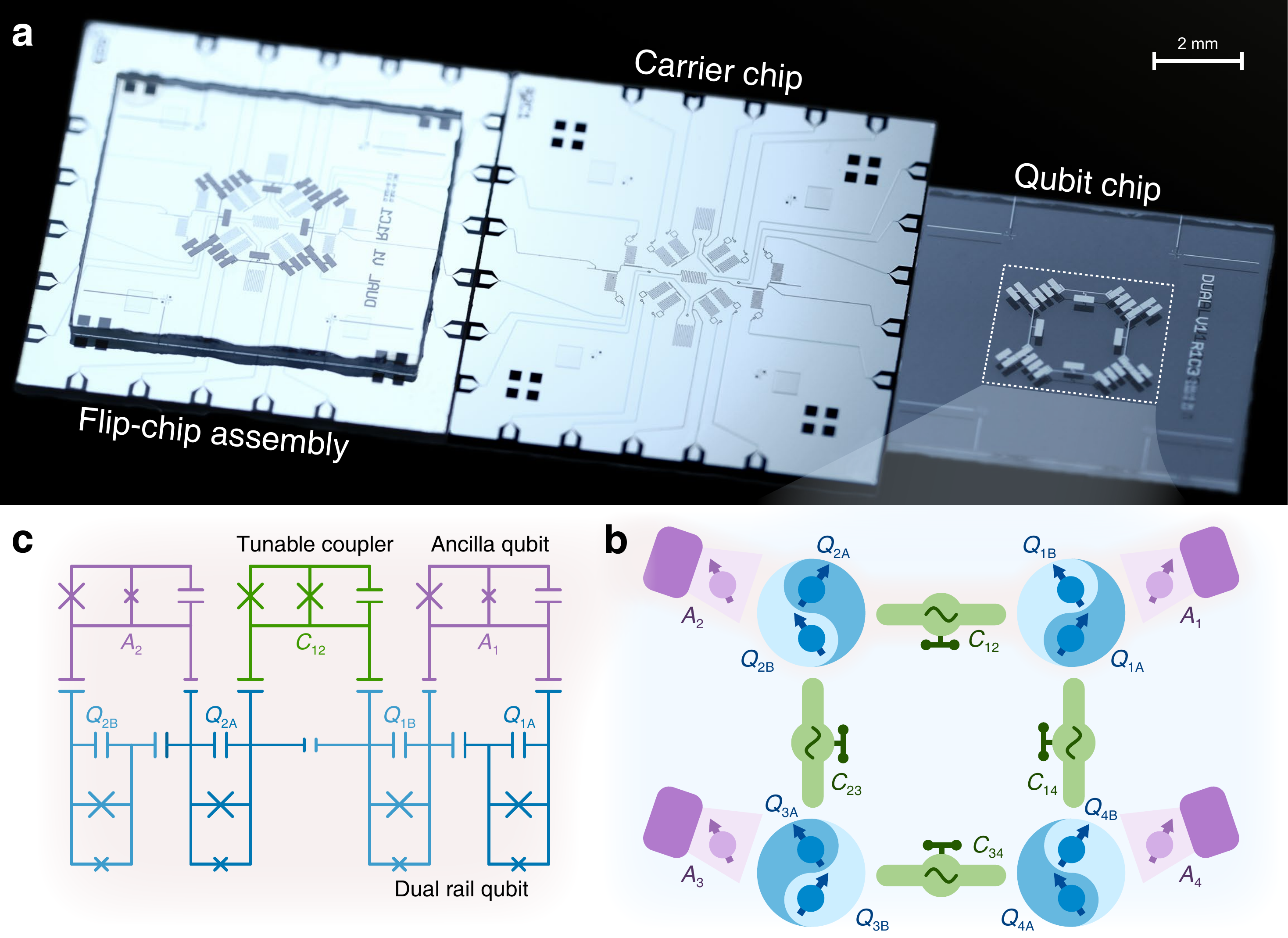}
    \caption{\label{fig1}
    \figtitle{Dual-rail superconducting quantum processor.}
    \subfiglabel{a}, Flip-chip assembled superconducting quantum processor (left), composited of a bottom carrier chip (middle) and a top qubit chip (right).
    \subfiglabel{b}, Illustration of the processor which consists of four logical qubits $Q_i$ ($i=1$, $
    \cdots$, 4) tunably coupled to each other through a coupler $C_{ij}$.
Each logical qubit consists of two physical transmons $Q_{iA}$ and $Q_{iB}$, and an associated ancilla qubit $A_i$ for erasure detection.
    \subfiglabel{c}, Circuit diagram of two logical qubits, with their associated ancilla qubits as well as the tunable coupler connecting them.
    }
\end{figure*}

Quantum error correction (QEC) is indispensable for realizing practical quantum computation by protecting the fragile quantum states against decoherence~\cite{Shor1995,Calderbank1996,Knill1997,Terhal2015}. By operating physical qubits below a critical error threshold, QEC schemes can suppress logical error rates exponentially through the redundant encoding of multiple physical qubits into a single logical qubit. While recent demonstrations in superconducting circuits~\cite{Sivak2023,Ni2023,Chen2021,Reinhold2020,LachanceQuirion2024,Acharya2023,Acharya2024}, trapped ions~\cite{Egan2021,RyanAnderson2021,Postler2022,Erhard2021,Neeve2022}, and neutral atoms~\cite{Graham2022,Evered2023,Bluvstein2023,Xu2024} approach the error thresholds of surface codes, practical implementation remains constrained by the quadratic resource overhead needed for exponential error suppression. This challenge has motivated two complementary strategies: engineering noise-biased physical qubits~\cite{Aliferis2008,Li2019,Guillaud2019,Xu2023,Puri2020,Mirrahimi2014} and exploiting detectable error channels through erasure conversion.


Erasures—errors with known locations—offer a unique opportunity to enhance QEC performance, as error correction protocols incorporating erasure conversion exhibit higher thresholds than conventional methods~\cite{Grassl1997,Bennett1997,Wu2022,Kubica2023,Sahay2023}. Harnessing this potential requires physical platforms that inherently produce detectable erasures, such as systems employing metastable states in neutral atoms and trapped ions~\cite{Wu2022,Ma2023,Kang2023}, or those utilizing dual-rail encoding in superconducting circuits. In the dual-rail scheme, the quantum states are encoded across qubit pairs of transmons~\cite{Shim2016,Campbell2020,Kubica2023,Levine2024} or cavity modes~\cite{Teoh2023,Koottandavida2024,Chou2024,Graaf2025}. By assigning the logical states in the single-excitation manifold of the qubit pairs, the amplitude damping ($T_1$) errors can be converted as erasures into leakage to non-computational states that can be detected with mid-circuit measurements. Moreover, in transmon-based dual-rail architectures, strong inter-qubit coupling not only facilitates this erasure conversion but also significantly suppresses residual dephasing errors~\cite{Campbell2020,Levine2024}, akin to the benefits of passive dynamical decoupling.


Recent experiments have validated key components of this paradigm. Dual-rail qubits have achieved millisecond-scale coherence with single-qubit gate errors below $10^{-4}$, complemented by high-fidelity erasure detection~\cite{Levine2024,Teoh2023,Chou2024,Koottandavida2024,Graaf2025}. However, the generation of multi-qubit logical entanglement---a prerequisite for both QEC and scalable quantum algorithms---has remained unrealized. This capability requires overcoming several key challenges including engineering tunable couplings between protected logical subspaces, converting errors during multi-qubit operations into erasures, and integrating ancilla-based erasure detection without compromising gate fidelities.

In this work, we establish a comprehensive framework for logical multi-qubit entanglement using dual-rail superconducting transmons. Our architecture leverages flip-chip packaging to enable scalable integration of four tunably coupled dual-rail qubits, each comprising transmon pairs that encode protected logical states. By directly modulating the flux of the physical qubits to drive the dual-rail qubits, we demonstrate high-fidelity single-qubit operations with error rates on the level of $10^{-5}$ and logical coherence times approaching $1\ \mathrm{ms}$---more than one order of magnitude improvement over physical qubit baselines. These advances are enabled by the static coupling between the transmon pairs that simultaneously suppresses dephasing noise and converts $T_1$ errors into detectable erasures. We further realize programmable entanglement through adiabatic coupler tuning, generating logical Bell states with $98.8\%$ fidelity and three-qubit GHZ states with $93.5\%$ fidelity. Notably, the logical entanglement can be maintained exceeding $100~\mu s$ with Bell state fidelity higher than 70\% in the absence of active error correction, underscoring the intrinsic error suppression afforded by dual-rail encoding. The measured $96.2\%$ process fidelity of our logical CNOT gate---currently limited by coupler induced decoherence---identifies promising pathways to optimize the gate performance toward fault-tolerant thresholds through improved coupler designs and refined pulse sequences.

\begin{figure*}[!t]
    \centering
    \includegraphics[width=0.95\textwidth]{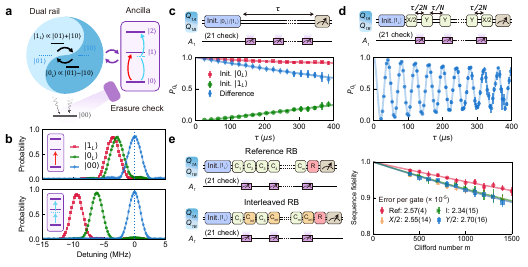}
    \caption{\label{fig2}
    \figtitle{Erasure check and single qubit operations.}
    \subfiglabel{a},
Schematic for single qubit operation and erasure check of a dual-rail qubit.
    \subfiglabel{b},
Conditional ancilla spectra with the dual-rail qubit in $\ket{0_L}$, $\ket{1_L}$, and $\ket{00}$, respectively. The top (bottom) panel shows the spectrum by driving the single-photon (two-photon) transition of the ancilla.
    \subfiglabel{c},
   $T_1$ relaxation of the dual-rail qubit with initial state $\ket{0_L}$ and $\ket{1_L}$, respectively. The population difference of the two is fit with an exponential decay to extract the logical relaxation time $T_1 = 0.98(1)$~ms.
   \subfiglabel{d},
   $T_2$ of the dual-rail qubit with CPMG($N$=256) sequence. The results are fit with an exponential sinusoid decay to extract $T_2 = 0.66(4)$~ms.
   \subfiglabel{e},
Randomized benchmarking of single-qubit logical gate. The experimental sequence is shown in the left panel. Each gate in the reference sequence is composed of $\pi/2$ pulses, with a single Clifford gate averaging 2.25 $\pi/2$ pulses. For the notation in right panel, the reference gate error represents the average error per $\pi/2$ pulse. The labels $I$, $X/2$, and $Y/2$ correspond to the gate errors for an idle pulse, a logical $X/2$ pulse, and a logical $Y/2$ pulse, respectively, all with a pulse duration of 25~ns. The erasure error per $\pi/2$ pulse, primarily due to leakage into the $|00\rangle$ state, is measured to be $6.4(1) \times 10^{-4}$ (data shown in Supplementary Information), more than one order of magnitude higher than the logical gate errors. This erasure error is effectively mitigated by erasure check, thereby preserving the fidelity of logical gate operations.
}
\end{figure*}

\section{Dual-rail processor}

The dual-rail superconducting quantum processor consists of a qubit chip and a carrier chip, assembled face-to-face using a non-galvanic flip-chip assembly technique~\cite{Satzinger2018,Satzinger2019Simple}, with the device picture shown in \rfig{fig1}{\bf a}. The qubit chip hosts four dual-rail qubits $Q_i$ ($i=1$, $\cdots$, 4) connected as a square (see the schematic in \rfig{fig1}{\bf b}), whereas the carrier chip hosts the Purcell filters, control and readout wiring circuitries~\cite{Qiu2025,Yang2024,Niu2023}.
The control wiring channels for $Q_{4A}$ is broken, therefore only the first three logical qubits are available on the processor. Each logical qubit consists of two transmon qubits $Q_{iA}$ and $Q_{iB}$~\cite{Koch2007} capacitively coupled to each other. The dual-rail logical qubit is defined in the single-excitation manifold of the physical qubit pair, where the logical subspace consists of the hybridized symmetric and antisymmetric states $|0_L\rangle = (|01\rangle - |10\rangle)/\sqrt{2}$ and $|1_L\rangle = (|01\rangle + |10\rangle)/\sqrt{2}$, with $|0\rangle$ and $|1\rangle$ being the ground and first excited states of the physical qubits respectively. The dual-rail encoding allows $T_1$ decay of the underlying physical qubits to be converted into erasure errors in the form of leakage to the $|00\rangle$ state. An ancilla qubit $A_i$ is capacitively coupled to the physical qubits for mid-circuit detection of the leakage to $|00\rangle$. The readout resonators of the four ancilla qubits are coupled to a common Purcell filter and share the same readout channel, facilitating the mid-circuit measurement of all the logical qubits simultaneously. The dual-rail qubits $Q_i$ and $Q_j$ are coupled to each other by implementing a tunable coupler~\cite{Yan2018,Xu2020,Sung2021} between the physical qubit $Q_{iA}$ and $Q_{jB}$, as shown by the circuit diagram in \rfig{fig1}{\bf c}. By tuning the frequency of the coupler, this could induce an effective $XX$ coupling in the logical subspace and mediate a $\sqrt{\rm{iSWAP}}$ entangling gate between the two logical qubits, based on which we can synthesize a logical controlled-NOT (CNOT) gate. More details about the device and experimental setup are available in the Supplementary Information.

\section{Erasure check and single qubit operations}

\begin{figure*}[!t]
    \centering
    \includegraphics[width=0.9\textwidth]{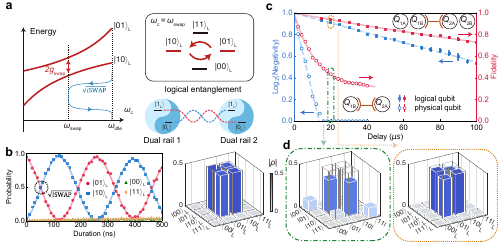}
    \caption{\label{fig3}
    \figtitle{Tunable coupling and robust entanglement.}
    \subfiglabel{a},
    Protocol to realize energy swap in the logical subspace of two coupled dual-rail qubits by tuning the coupler frequency $\omega_c$. While idling, the logical states $|01\rangle_L$ and $|10\rangle_L$ are non-degenerate. By tuning down the coupler frequency to $\omega_{\mathrm{swap}}$, the two logical states become degenerate with a coupling strength $g_{\mathrm{swap}}$. The right panel shows the energy diagram during the logical swap process and a cartoon schematic.  
    \subfiglabel{b},
    Population of each logical state as a function of the swap duration with initial state $|01\rangle_L$. Solid lines are sinusoidal fit. The dotted circle marks the pulse duration of a $\sqrt{i\mathrm{SWAP}}$ gate. The right panel shows the density matrix of the logical Bell state generated by a single logical $\sqrt{i\mathrm{SWAP}}$ gate, where the colored bars are obtained from QST and the gray frames correspond to the ideal value. 
    \subfiglabel{c},
    Decay of state fidelity and logarithm negativity as a function of the delay time after the generation of a Bell state using two logical qubits and physical qubits respectively. CPMG pulses and mid-circuit erasure checks are used during the delay time. The solid lines are exponential fit for the fidelity data. The dashed lines are linear fit for the logarithm negativity data.
    \subfiglabel{d},
    The density matrix of the states with a 20~$\mu s$ delay time after Bell state generation for the logical qubits (right panel) and physical qubits (left panel) respectively. The corresponding data points in \textbf{(c)} are highlighted by orange dotted line and green dotted dash line. 
    }
\end{figure*}

We begin by calibrating single-qubit operations and the erasure check for each dual-rail qubit. In Fig.~\ref{fig2}\textbf{a}, we show the schematic for the operations of a single dual-rail qubit. The two physical qubits are brought into resonance to establish the eigenstates of the dual-rail qubit, $|0\rangle_L$ and $|1\rangle_L$. Owing to the coupling between the physical qubits, these states exhibit an energy gap of 150–180~MHz, contingent on the qubit frequencies. Coherent rotations between $|0\rangle_L$ and $|1\rangle_L$ are achieved through modulating the flux of either physical qubit. By precisely tuning the frequency, amplitude, and phase of the driving pulse, high-fidelity logical single-qubit gates can be implemented.

During the experimental sequence, mid-circuit erasure checks are performed to detect the presence of erasure errors. This check is facilitated by an ancilla qubit coupled to the dual-rail qubit in the dispersive regime. The erasure check consists of a conditional $\pi$ pulse applied to the ancilla, which is resonant only when the dual-rail qubit is in the $|00\rangle$ state, followed by state readout of the ancilla. The dispersive coupling between the ancilla and the dual-rail qubit results in different ancilla frequencies depending on the state of the dual-rail qubit. The frequency of the conditional $\pi$ pulse is chosen such that the ancilla is only excited if the dual-rail qubit is in the $|00\rangle$ state (indicating an erasure error) and remains in the ground state if the dual-rail qubit is in the logical space (indicating no erasure error). The driving pulse can consist of either a single-photon excitation, which places the ancilla in the $|1\rangle$ state, or a two-photon excitation, which places the ancilla in the $|2\rangle$ state. The corresponding ancilla spectra of these two schemes with dual-rail qubit in different states are shown in Fig.~\ref{fig2}\textbf{b}. Using the two-photon excitation method yields a larger dispersive shift and improved readout fidelity, thus this method is employed for erasure checks in the subsequent experiments, unless otherwise noted.

\begin{figure*}[!t]
    \centering
    \includegraphics[width=0.8\textwidth]{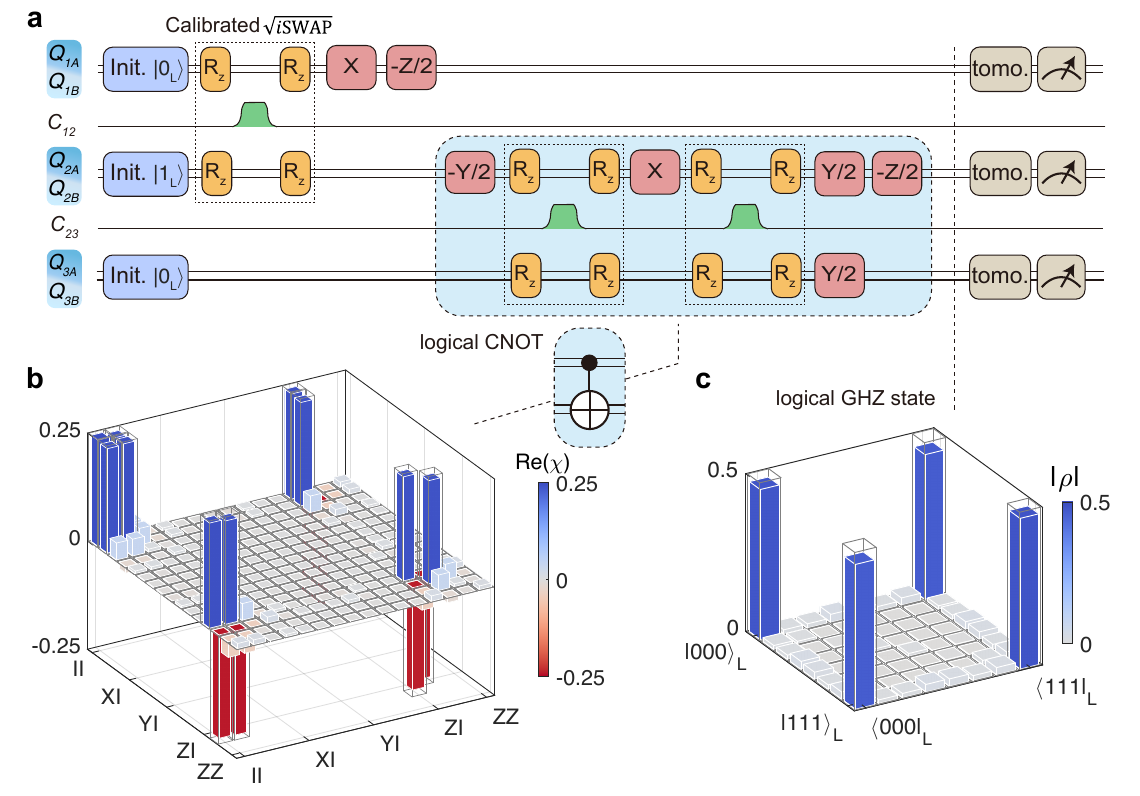}
    \caption{\label{fig4}
 \figtitle{CNOT gate synthesis and multi-qubit entanglement.}
 \subfiglabel{a},
Pulse sequence for generating a three-qubit logical GHZ state. The four $R_z$ gates correct the accumulated single qubit phases and the relative phase due to qubit frequency difference during the logical $\sqrt{i\mathrm{SWAP}}$ gate. The gates in the blue dashed frame forms a logical CNOT gate. 
 \subfiglabel{b},
QPT of the logical CNOT gate, with a process fidelity of 96.2(3)\%. Colored bars are the measured real parts of the $\chi$ matrix and the grey frame corresponds to the ideal value. All elements of the imaginary parts are less than 0.05 (results shown in Supplementary Information).
 \subfiglabel{c}
QST of the logical GHZ state, with a state fidelity of 93.9(2)\%. Colored bars are the measured density matrix and the grey frame corresponds to the ideal value.
}
\end{figure*}

To evaluate error rates of single-qubit operations within the logical subspace, we perform coherence measurements and randomized benchmarking experiments~\cite{Magesan2011,Magesan2012}, with postselection on the shots where no erasure errors occur, as shown in Fig.~\ref{fig2}\textbf{c-e}. The measured coherence times are approximately an order of magnitude longer than those of the corresponding physical qubits, with a logical $T_1 = 0.98$~ms and a logical CPMG $T_2 = 0.66$~ms. The average fidelity of a single $\pi/2$ gate is $99.9974\%$, approaching the coherence-limited fidelity in the logical subspace.
Notably, the resonant coupling between $|01\rangle$ and $|10\rangle$ with strength $g$ acts as a passive decoupling mechanism which strongly suppresses the impact of low-frequency noise on the underlying physical qubits~\cite{Campbell2020}, analogous to continuous dynamical decoupling in driven systems~\cite{Yan2013,Guo2018}.
This effect is reflected in the dual-rail energy gap $E_{DR} \approx \sqrt{(2g)^2 + \delta^2}$, where $\delta$ is the frequency difference between the two physical qubits due to inherent frequency noise. With $g \gg \delta$, the frequency noise is suppressed, with $E_{DR} \approx 2g + \delta^2/4g$, leading to a dual-rail energy gap that is orders of magnitude more stable than the underlying physical qubits. This noise suppression results in a large erasure noise bias, significantly reducing dephasing errors within the logical subspace.

\section{Tunable coupling and robust entanglement}


Next, we demonstrate the logical two-qubit gate for dual-rail encoding. Consider a system consisting of two dual-rail qubits, each containing two physical qubits: $Q_{\mathrm{1A}}$, $Q_{\mathrm{1B}}$, $Q_{\mathrm{2A}}$, and $Q_{\mathrm{2B}}$, where 1 and 2 denotes different dual-rail qubits. Given a tunable physical $XX$ coupling between $Q_{\mathrm{1B}}$ and $Q_{\mathrm{2A}}$, it has been proposed that a logical $\sqrt{i\mathrm{SWAP}}$ gate can be implemented within the dual-rail logical subspace~\cite{Kubica2023}. To achieve a high-fidelity logical $\sqrt{i\mathrm{SWAP}}$ gate, two conditions must be met. First, the logical frequencies of the two dual-rail qubits should be identical to enable energy exchange. Second, the frequency detuning of physical qubits in the two dual-rail qubits should be large compared to the coupling strength between different physical qubits, and the qubits' nonlinearity. With these conditions, high-fidelity logical $\sqrt{i\mathrm{SWAP}}$ gate could be achieved by adiabatically tuning the physical coupling strength.

To realize this in the experiment, a tunable coupler between $Q_{\mathrm{1B}}$ and $Q_{\mathrm{2A}}$ is used to effectively tune the coupling strength dynamically. Due to the hybridization of the coupler and the dual-rail qubits, tuning the coupler would inevitably shift the qubit frequency and also induce state leakage outside the logical subspace. Considering this, while the coupler is at the idling frequency $\omega_{\mathrm{idle}}$, the two dual-rail qubits have different logical frequency and weak $XX$ interaction so energy swap is not activated. Specifically, $\omega_{\mathrm{idle}}$ is set at the frequency where the logical $ZZ$ interaction of the two dual-rail qubits is zero to prevent accumulation of residual $ZZ$ phase. To enable the energy swap, the coupler is adiabatically tuned to the frequency $\omega_{\mathrm{swap}}$, where logical states $|01\rangle_L$ and $|10\rangle_L$ are on resonance, as illustrated by Fig.~\ref{fig3}\textbf{a} (see Supplementary Information for experimental spectral data). While on resonance, the $XX$ interaction couples the logical states $|10\rangle_L$ and $|01\rangle_L$ by a second-order process through states outside the logical subspace such as $|1100\rangle$, $|0200\rangle$ and etc. As a result, the swap process inevitably causes some leakages; however, these leakages can be detected as erasure errors and do not limit the gate fidelity. Figure~\ref{fig3}\textbf{b} shows a clear population swap between states $|01\rangle_L$ and $|10\rangle_L$ while the coupler is tuned to frequency $\omega_{\mathrm{swap}}$. 

To further confirm a Bell state is generated by the $\sqrt{i\mathrm{SWAP}}$ gate, we use quantum state tomography (QST)~\cite{Steffen2006} to reconstruct the density matrix of the logical entangled state, as shown in the right panel of Fig.~\ref{fig3}\textbf{b}. The corresponding state fidelity of the generated logical Bell state is $98.8(1)$\%. Furthermore, we confirm the generated entanglement in the logical subspace is also protected by the dual-rail encoding. Figure~\ref{fig3}\textbf{c} shows the comparison of entanglement degradation for a Bell state generated by two physical qubits and by two dual-rail qubits, respectively. The data are obtained with variable delay time between the Bell state generation and the QST measurements. Both the state fidelity and logarithmic negativity data show that the entanglement of the logical qubit is preserved for approximately one order of magnitude longer, unambiguously indicating the protection by dual-rail encoding. To illustrate it more clearly, Fig.~\ref{fig3}\textbf{d} shows the reconstructed density matrix of the state at 20~$\mu s$ delay after thed Bell state generation for both the logical qubit (right panel) and the physical qubit (left panel). It is evident that while the physical Bell state almost completely loses the entanglement information, the logical Bell state still maintains a high fidelity of $93.5(4)$\%. Notably, the logical Bell state's fidelity remains above 70\% after a delay exceeding 100~$\mu s$, demonstrating that dual-rail encoding effectively mitigates decoherence errors in the encoded qubits.

The millisecond-scale coherence time of the dual-rail qubits suggest a coherence limited fidelity higher than 99.9\%~\cite{Sete2024}. However, the obtained fidelity of the logical Bell state is much less than that. We attribute the infidelity to coupler induced decoherence. Even though the dual rail qubit have long coherence while the coupler is idling, the coherence dramatically decreases while the coupler is tuned to $\omega_{\mathrm{swap}}$ (see Supplementary Information for details). This originates from strong hybridization of the energy levels between the qubits and the coupler as the coupler frequency is close to the physical qubit. This leads to extra decoherence channels in the logical subspace and degrade the fidelity of the generated Bell state.

\section{Logical multi-qubit entanglement}

Next, we demonstrate the scalability of the dual-rail system by extending entanglement generation to multiple qubits. We begin by demonstrating the generation of a logical CNOT gate using two $\sqrt{i\mathrm{SWAP}}$ gates and additional logical single-qubit gates, as shown in Fig.\ref{fig4}\textbf{a}. It is noteworthy that the generated $\sqrt{i\mathrm{SWAP}}$ gate induces both single-qubit and $ZZ$ phases in the logical subspace, as tuning the coupler to $\omega_{\mathrm{swap}}$ shifts all qubit frequencies and the logical $ZZ$ interaction becomes nonzero. The single-qubit phase can be compensated using logical single-qubit gates with a calibration routine (see Supplementary Information for details), while the logical $ZZ$ phase cannot be easily cancelled. To mitigate the effect of the $ZZ$ phase, a logical $\pi$ pulse is inserted between the two $\sqrt{i\mathrm{SWAP}}$ gates, along with additional single-qubit gates, resulting in a $ZZ$-free logical CNOT gate. To verify this, we perform quantum process tomography (QPT)~\cite{Mohseni2007,Neeley2008Process} in the logical subspace to extract the process matrix. Figure~\ref{fig4}\textbf{b} shows the real part of the process matrix $\mathrm{Re}(\chi)$, corresponding to a process fidelity of 96.2(3)\% for the CNOT gate.

The logical CNOT gate, in combination with logical single-qubit operations, constitutes a universal gate set--an essential milestone for scaling the dual-rail system. We showcase this scalability by generating a three-qubit logical GHZ state, with the pulse sequence shown in Fig.~\ref{fig4}\textbf{a}. First, the three dual-rail qubits are initialized to $|0\rangle_L$, $|1\rangle_L$, and $|0\rangle_L$, respectively. A logical Bell state, $(|00\rangle_L + |11\rangle_L)/\sqrt{2}$, is then generated using a calibrated $\sqrt{i\mathrm{SWAP}}$ gate and two single-qubit gates. Next, a logical CNOT gate is applied between the second and third dual-rail qubits, generating the three-qubit logical GHZ state, $(|000\rangle_L + |111\rangle_L)/\sqrt{2}$. QST in the logical subspace is performed to extract the density matrix of the generated logical state, as shown in Fig.~\ref{fig4}\textbf{d}, with a state fidelity of 93.9(2)\%. 

Similar to the infidelity observed in the generated logical Bell state, infidelity in both the logical CNOT gate and the generated logical GHZ state could be attributed to the additional decoherence induced by the coupler while tuned to $\omega_{\mathrm{swap}}$. It is worth mentioning that the surface code can be realized with dynamic quantum circuits using $i$SWAP gates instead of the traditional CNOT, which could also be generated in our protocol, extending the set of viable gates for error correction without additional overhead~\cite{Google2024dynamic}.

\section{Discussion and outlook}

In summary, we have devised and demonstrated what is, to our knowledge, the first entangling gate for superconducting erasure qubits. In this system, the lifetime of each logical qubit is on the millisecond scale, and the logical single-qubit gate fidelity is as low as $2.6\times10^{-5}$. Logical entanglement is achieved by mediating the flux of the tunable coupler between dual-rail qubits to generate a logical $\sqrt{i\mathrm{SWAP}}$ gate. A logical Bell state with a fidelity of 98.8\% is generated, with an entanglement storage time an order of magnitude longer than that of the physical counterpart. Furthermore, the logical operation can be easily scaled up by constructing a universal gate set with the logical CNOT gate, which currently has a gate fidelity of 96.2\%. A three-qubit GHZ state with a fidelity of 93.9\% is generated to showcase the scalability of the protocol.

Besides the applications in QEC, the dual-rail system, with its ability to generate entangled states with significantly longer coherence times than physical qubits, offers promising applications in quantum networks~\cite{Wehner2018,Zhong2021,Niu2023} and quantum metrology~\cite{Toth2014,Degen2017,Niroula2024,Li2024}. Although the current logical CNOT gate fidelity is suboptimal, the millisecond-scale coherence of the dual-rail qubit suggests that the coherence-limited fidelity could be elevated above 99.9\%. Future work will concentrate on refining coupler architectures and optimizing entanglement pulse sequences to improve the logical two-qubit gate performance.


Note added--While preparing the manuscript, we became
aware of a relevant experiment~\cite{Mehta2025}, which also demonstrates the two-qubit entangling gate for dual-rail erasure qubits, but is based on 3D superconducting cavities.

\subsection*{Acknowledgements}
This work was supported by the Science, Technology and Innovation Commission of Shenzhen Municipality (KQTD20210811090049034), the Innovation Program for Quantum Science and Technology (2021ZD0301703),  the National Natural Science Foundation of China (12174178, 12404582). Shenzhen Science and Technology Program (Grant No. RCBS20231211090815032).

\subsection*{Author contributions}
Y.Z. conceived the experiment and supervised the project.
W.H. designed and tested the device.
X.D.S. developed the FPGA program for the custom electronics built by J.Z.
X.L. performed the measurement and analyzed the data.
All authors contributed to the experimental setup, discussions of the results and writing of the manuscript.

\subsection*{Competing interests}
The authors declare no competing interests.

\subsection*{Data availability}
The data that support the plots within this paper and other findings of this study are available from the corresponding authors upon request.

\bibliography{ref_main,dual_rail}
\bibliographystyle{naturemag}

\end{document}


\title{Supplementary Information for ``Logical multi-qubit entanglement with dual-rail superconducting qubits''}


\maketitle

\setcounter{equation}{0}
\setcounter{figure}{0}
\setcounter{table}{0}
\setcounter{page}{1}

\renewcommand{\theequation}{S\arabic{equation}}
\renewcommand{\thefigure}{S\arabic{figure}}
\renewcommand{\thetable}{S\arabic{table}}

\newcommand{\figtitle}[1]{\textbf{#1}}
\newcommand{\subfiglabel}[1]{\textbf{#1}}

\tableofcontents

\clearpage

\section{Experimental setup and device information}\label{sec:setup}

\begin{figure}[b!]
    \centering
    \includegraphics[width=0.9\linewidth]{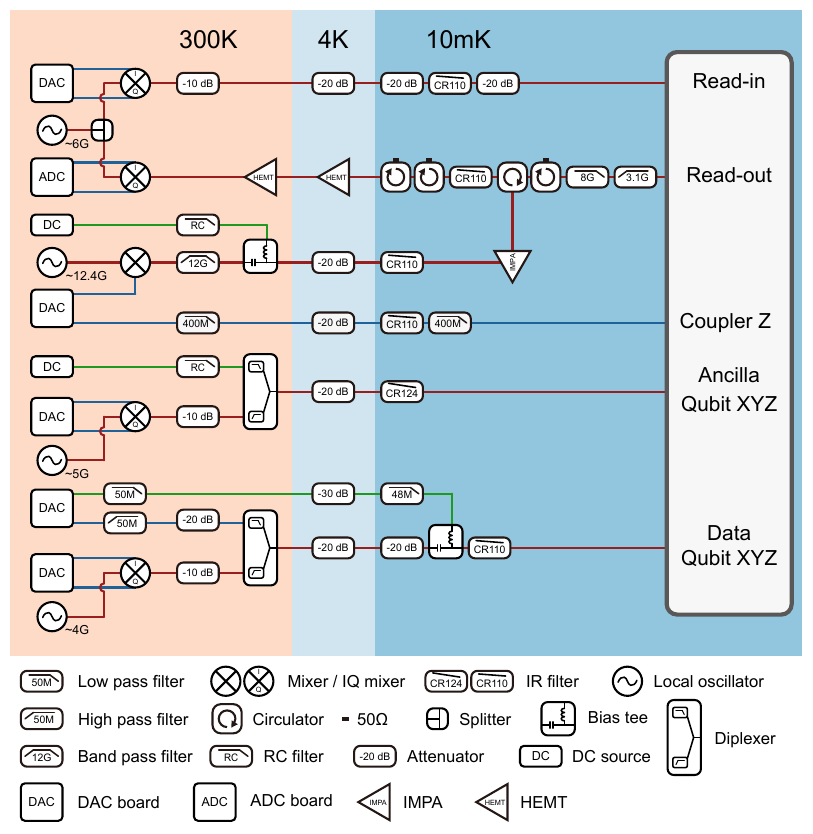}
    \caption{\textbf{Schematic diagram of the experimental setup.} Wiring of different colors corresponds to distinct signal frequencies. The flux bias ($<$40~MHz) for the qubit and IMPA is transmitted through the green signal lines, while the blue lines are dedicated to the fast bias and dual-rail qubit driving signals (100-200~MHz). 
    Microwave signals ($>$1 GHz), including those for transmon qubit driving, readout, and IMPA pumping, are routed via the red lines.}
    \label{FigSI_dual_setup}
\end{figure}

The experimental setup for qubit control and readout, including cryogenic wiring and room-temperature microwave electronics, is schematically illustrated in Fig.~\ref{FigSI_dual_setup}. Our quantum processor, comprising eight data qubits, four ancilla qubits, four tunable couplers, and three readout lines, is thermally anchored to the mixing chamber stage (10~mK) of a Bluefors LD400 dilution refrigerator. The readout architecture employs frequency multiplexing, allowing each readout line to simultaneously interrogate four qubits through their respective readout resonators.

For qubit control and measurement, we utilize custom-designed digital-to-analog converters (DACs, 2~Gs/s sampling rate) and analog-to-digital converters (ADCs, 1~Gs/s sampling rate)~\cite{zhang2024}. Single-qubit XY control and dispersive readout microwave signals are generated through IQ mixing, where DAC outputs are up-converted using an external microwave signal generator as the local oscillator. Flux control signals, including slow bias, fast bias, and dual-rail driving signals, are directly synthesized by the DACs. Notably, for dual-rail data qubit control, we implement a low-temperature signal combination scheme: the slow bias signals and fast signals (dual-rail driving signal and single-qubit XY control) are combined using a bias tee at the mixing chamber stage. Compared to room-temperature signal combination, this configuration reduces thermal noise and enhances coherence time within the dual-rail logical space. All control lines incorporate strategically placed filters and attenuators at different temperature stages for thermalization and noise suppression.

The readout chain features a multi-stage amplification scheme. At the preamplifier stage, we employ a custom-designed impedance-matched Josephson parametric amplifier (IMPA), followed by a high electron mobility transistor (HEMT) amplifier at the 4~K stage, with additional room-temperature amplification. Circulators and different filters are installed in mixing chamber stage to block noise from higher-temperature stages and prevent backflow of noise from the IMPA into the quantum chip, thereby minimizing its impact on qubit coherence. The pump tone of the IMPA is additionally modulated using a mixer at room temperature, ensuring IMPA activation only during readout pulses to minimize potential decoherence effects.



\begin{figure}[b!]
    \centering
    \includegraphics[width=0.98\linewidth]{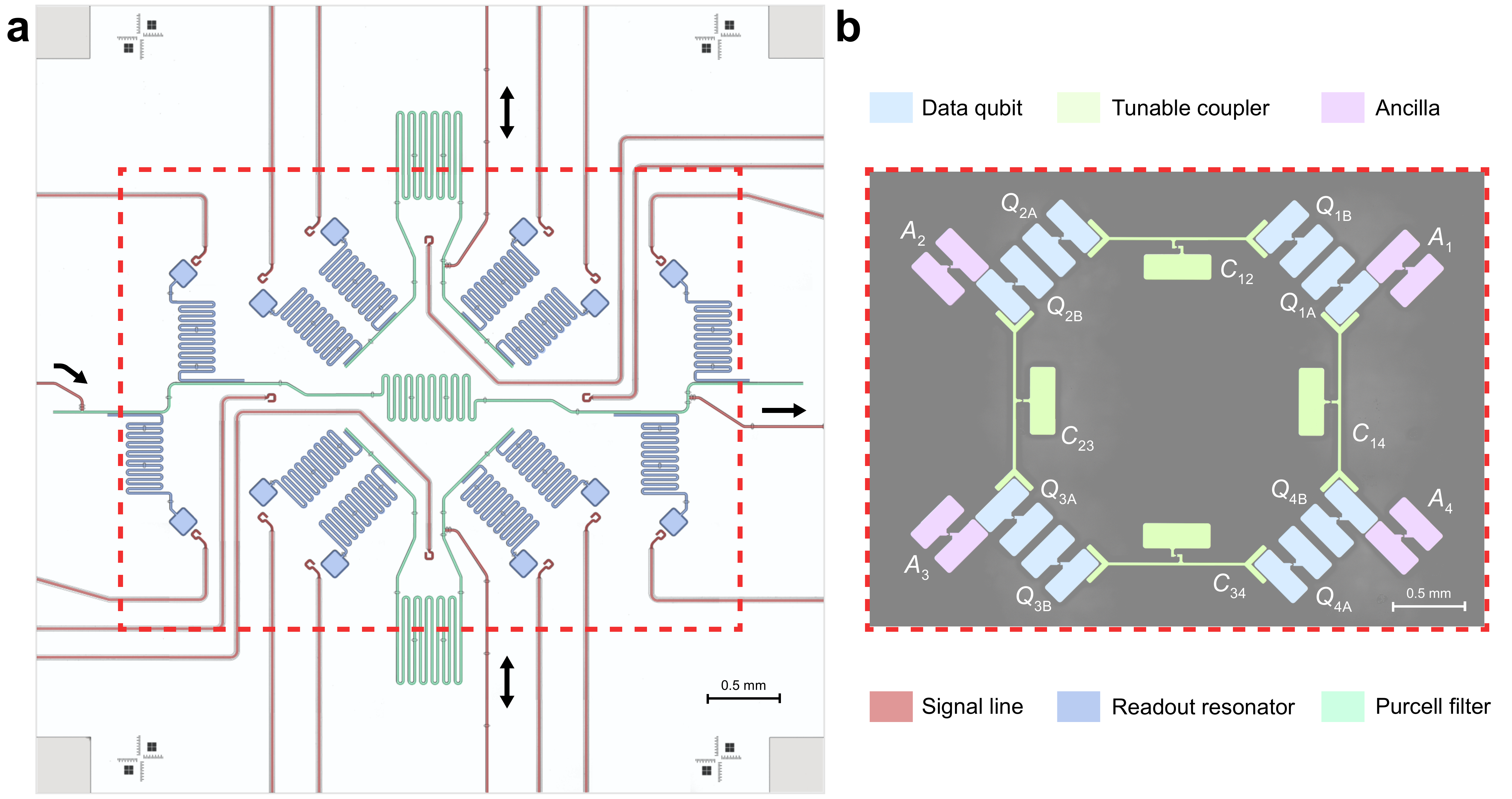}
    \caption{\textbf{False-color micrograph of of the quantum processor.} \subfiglabel{a}, The carrier chip for qubit control and readout. The readout resonators (blue) of the ancilla qubits share the same Purcell filter in the center, with the input and output port located on the left and right respectively. The four data qubits on the top share the same Purcell filter, while the bottom four data qubits share another Purcell filter respectively.
    These two Purcell filters have only one port serving as both input and output decoupled by a circulator. \subfiglabel{b}, The qubit chip comprising data qubits (light blue), ancilla qubits (pink), and tunable couplers (green).
    The red dashed rectangle in {\bf a} marks the position where the patterns in {\bf b} is located when flip-chip bonded.}
    \label{FigSI_dual_chip}
\end{figure}


The micrograph picture of the quantum processor is shown in Fig.~\ref{FigSI_dual_chip}. Basic parameters for the data qubits and ancillas are summarized in Table~\ref{Table1}. The specific qubit parameters for experimental results shown in the main text Fig.~2-4 are summarized in Table~\ref{Table2}-\ref{Table4}.

\begin{figure}[b!]
    \centering
    \includegraphics[width=0.8\linewidth]{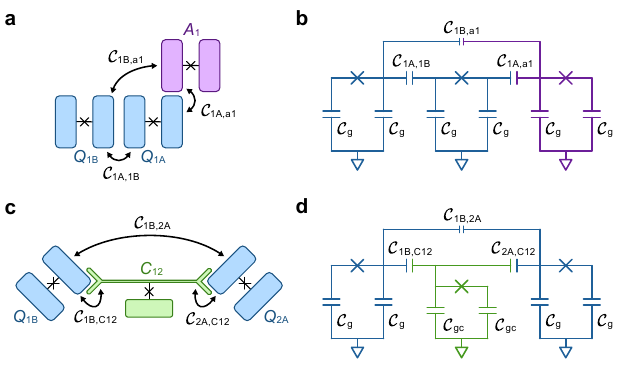}
    \caption{\textbf{Schematic diagraman of coupling configuration.} \subfiglabel{a}, Sketch of two data qubits~(blue) and ancilla~(purple) with DDA configuration,. \subfiglabel{b}, Circuit diagram of DDA structure. \subfiglabel{c}, Sketch of two data qubits~(blue) and coupler~(green) with DCD configuration,. \subfiglabel{d}, Circuit diagram of DCD structure. The arrow indicates the primary coupling capacitance in the schematic diagram.}
    \label{FigSI_dual_circuit}
\end{figure}

Our quantum processor utilizes floating transmons for data qubits, ancilla qubits, along with tunable couplers, a widely adopted design in two-dimensional superconducting chip.
In our flip-chip architecture, the pad-pad capacitance of floating qubit is substantially smaller than its pad-ground capacitance, resulting in qubit capacitance dominated by the latter.
The coupling geometries in qubit chip comprises two configurations: (1) the Data-Ancilla (DDA) configuration (Fig.~\ref{FigSI_dual_circuit}~\textbf{a}) and (2) the Data-Coupler (DCD) configuration (Fig.~\ref{FigSI_dual_circuit}~\textbf{c}). 
The effective circuits, retaining the Josephson junctions and dominant capacitive elements while neglecting minor pad-pad and stray capacitance, are shown in Fig.~\ref{FigSI_dual_circuit}~\textbf{b}~(DDA) and Fig.~\ref{FigSI_dual_circuit}~\textbf{d}~(DCD). 
The pad-to-ground capacitance is characterized by $\mathcal{C}_{g}$ for all data and ancilla qubits and $\mathcal{C}_{gc}$ for both pads of the floating coupler, with each capacitance type maintaining uniform values across its respective components.

In the DDA configuration, the indirect capacitive coupling through the network is negligible due to minimal pad-pad capacitance in $Q_{1A}$. 
The coupling strength between the data qubit $Q_{1B}$ and the ancilla $A_{1}$ follows 
\begin{equation}
    g_{1B,a1} \approx \frac{\sqrt{\omega_{1B}\omega_{a1}}}{8} \frac{\mathcal{C}_{1B,a1}}{\mathcal{C}_g/2}, 
    \label{eq_coupling_gda}
\end{equation}
with analogous expressions governing the other two coupling strengths.
For our specific design parameters at the frequency $4$~GHz, the coupling strengths are $g_{1B,a1}/2\pi = 3.65$~MHz , $g_{1A,1B}/2\pi = 82$~MHz and $g_{1A,a1}/2\pi = -56.3$~MHz. 
The sign of each coupling term is determined by whether the connected qubits share the same pad or different pads in the circuit geometry.

In the DCD configuration, the coupling strengths between data qubit and coupler follow a similar form to Eq.~\ref{eq_coupling_gda}. 
However, when two data qubits couple to the same pad of the coupler, an additional indirect coupling via network emerges.
The total capacitive coupling between $Q_{1B}$ and $Q_{2A}$ is given by 
\begin{equation}
    g_{1B,2A} \approx \frac{\sqrt{\omega_{1B}\omega_{2A}}}{8}\left(
    \frac{\mathcal{C}_{1B,2A}}{\mathcal{C}_g/2} + \frac{\mathcal{C}_{1B,C12} \mathcal{C}_{2A,C12}}{\mathcal{C}_g \mathcal{C}_{gc}/4}
    \right), 
    \label{eq_coupling_gqq}
\end{equation}
where the second term represents the contribution from the indirect capacitive network.  
For our device parameters at the frequency $4$~GHz, the coupling strengths are $g_{1B,C12}/2\pi = g_{2A,C12}/2\pi = 101.2$~MHz and $g_{1B,2A}/2\pi = 12.8$~MHz. 
When the detuning coupler is effectively decoupled from the qubit system, the total qubit-qubit coupling is modified by the coupler-mediated interaction~\cite{yan2018tunable}. This effective coupling can be expressed as
\begin{equation}
    g_\text{eff} \approx g_{1B,2A}+\frac{g_{1B,C12}g_{2A,C12}}{2}\left(
    \frac{1}{\Delta_{1B,C12}}+\frac{1}{\Delta_{2A,C12}}
    -\frac{1}{\sum_{1B,C12}}-\frac{1}{\sum_{2A,C12}}
    \right),
    \label{eq_coupling_geff}
\end{equation}
where $\Delta_{k,C12}=\omega_k-\omega_{C12}$ and $\sum_{k,C12}=\omega_k+\omega_{C12}$ ($k = 1B,2A$) denote the detuning parameters. 
The effective qubit-qubit coupling strength $g_{\text{eff}}$, as described by Eq.~\ref{eq_coupling_geff}, exhibits continuous tunability via the coupler frequency $\omega_{C12}$


\begin{table}[!b]
\centering
\begin{tabular}{p{2.5cm} p{1.2cm}p{1.2cm}p{1.2cm}p{1.2cm}p{1.2cm}p{1.2cm}p{1.2cm}p{1.2cm}p{1.2cm}} 
\hline
\hline
 & $Q_{{1A}}$ & $Q_{{1B}}$ & $Q_{{2A}}$ & $Q_{{2B}}$ & $Q_{{3A}}$ & $Q_{{3B}}$ & $A_1$ & $A_2$ & $A_3$\\
\hline
$f_{01}^{\mathrm{max}}$ (GHz) & 5.17 & 5.17 & 5.14 & 5.29 & 5.23 & 5.31 & $\sim$6 & $\sim$6 & $\sim$6 \\
$f_{01}^{\mathrm{min}}$ (GHz) & 3.68 & 3.69 & 3.55 & 3.83 & 3.78 & 3.70& 4.05 & 4.08 & 4.08 \\
$\eta$/2$\pi$ (GHz) & $-0.20$ & $-0.20$ & $-0.22$ & $-0.21$ & $-0.21$ & $-0.20$ & $-0.19$ & $-0.2$ & $-0.19$ \\
\hline
\hline
\end{tabular}
\caption{Frequency range and anharmonicity of data qubits and ancillas.}
\label{Table1}
\end{table}

\begin{table}[h!]
\centering
\begin{tabular}{ c >{\centering\arraybackslash}p{2cm} >{\centering\arraybackslash}p{2cm} c} 
\hline
\hline
& $Q_{{1A}}$-$Q_{{1B}}$ &  &   $A_1$ (2-photon) \\
\hline 
$f_{|0_L\rangle}$ (GHz) & 4.367 & $f_{01}$ (GHz) & 4.978 \\
$T_{1,|0_L\rangle}$ ($\mu s$) & 63 & $T_1$ ($\mu s$) & 7 \\
$f_{|1_L\rangle}$ (GHz) & 4.541 & \parbox[h]{2.0cm}{\setlength{\baselineskip}{10pt} readout fid.\\$F_{|0\rangle}$} & 99.6\% \\
$T_{1,|1_L\rangle}$ ($\mu s$) & 56 & $F_{|1\rangle}$ & 97.3\% \\
$f_{L}$ (MHz) & 174.0 &  &  \\
$T_{1,L}$ (ms) & 0.98 &  & \\
$T_{2,L}$ (ms) & 0.66 & &  \\
\parbox[h]{2.6cm}{\setlength{\baselineskip}{10pt} readout fid.\\$F_{|0_L\rangle}$} & 97.5\% &  &  \\
$F_{|1_L\rangle}$ & 96.7\% & &  \\
$F_{|00\rangle}$ & 99.8\% & & \\
\parbox[h]{2.6cm}{\setlength{\baselineskip}{10pt} logical 1Q \\ gate fid.}  & 99.997\% & & \\
\hline
\hline
\end{tabular}
\caption{Parameters of dual-rail qubits and ancilla qubits for experimental results in main text Fig.~2. $f_{|0_L\rangle}$ ($f_{|1_L\rangle}$) is the frequency to drive the dual rail qubit from $|00\rangle$ to $|0_L\rangle$ ($|1_L\rangle$). $T_{1,|0_L\rangle}$ ($T_{1,|1_L\rangle}$) is the relaxation time from $|0_L\rangle$ ($|1_L\rangle$) to $|00\rangle$, close to the $T_1$ of individual physical qubit. $f_{L}$, $T_{1,L}$, and $T_{2,L}$ are the frequency, relaxation time, CPMG dephasing time of the dual-rail qubit in the logical space. The readout fidelity in the logical space, $F_{|0_L\rangle}$, $F_{|1_L\rangle}$, and $F_{|00\rangle}$, are obtained by a joint readout of the two physical qubits after adiabatically tuning the qubits off from the frequency degeneracy point thus the logical state $|0_L\rangle$ ($|1_L\rangle$) adiabatically changes to physical states $|10\rangle$ ($|01\rangle$).}
\label{Table2}
\end{table}

\begin{table}[h!]
\centering
\begin{tabular}{>{\centering\arraybackslash}p{2.1cm} >{\centering\arraybackslash}p{2cm} >{\centering\arraybackslash}p{2cm} ccccc} 
\hline
\hline
& $Q_{{1A}}$-$Q_{{1B}}$ & $Q_{{2A}}$-$Q_{{2B}}$ &  & $A_1$ (2-photon) & $A_2$ (2-photon) & $Q_{{1B}}$ & $Q_{{2A}}$ \\
\hline 
$f_{|0_L\rangle}$ (GHz) & 4.304 & 3.828 & $f_{01}$ (GHz) & 4.700 & 4.784 & 4.225 & 3.906 \\
$T_{1,|0_L\rangle}$~($\mu s$) & 58 & 86 & $T_1$ ($\mu s$) & 14 & 31 & 65 & 65 \\
$f_{|1_L\rangle}$~(GHz) & 4.475 & 3.980 & $T_2$ ($\mu s$) & - & - & 13 & 15 \\
$T_{1,|0_L\rangle}$~($\mu s$) & 35 & 64 & \parbox[h]{2.0cm}{\setlength{\baselineskip}{10pt} readout fid.\\$F_{|0\rangle}$} & 99.5\% & 98.0\% & 99.9\% & 99.5\% \\
$f_{L}$~(MHz) & 171.8 & 152.5 & $F_{|1\rangle}$ & 92.6\% & 97.3\% & 98.4\% & 97.8\% \\
$T_{1,L}$~(ms) & 0.59 & 1.29 & &  & &  &  \\
$T_{2,L}$~(ms) & 0.24 & 0.62 & &  & &  &  \\
\parbox[h]{2.0cm}{\setlength{\baselineskip}{10pt} readout fid.\\$F_{|0_L\rangle}$} & 96.6\% & 97.2\% & &  & &  &  \\
$F_{|1_L\rangle}$ & 96.2\% & 97.0\% &  &  &  &  &  \\
$F_{|00\rangle}$ & 99.8\% & 99.6\% & &  & &  &  \\
\parbox[h]{2.6cm}{\setlength{\baselineskip}{10pt} logical 1Q \\ gate fid.}   & 99.993\% & 99.996\% & &  & &  & \\
\hline
\hline
\end{tabular}
\caption{Parameters of dual-rail qubits, ancilla qubits and physical qubits for experimental results in main text Fig.~3. The last two columns record the qubit parameters used to generate the physical Bell state.}
\label{Table3}
\end{table}

\begin{table}[h!]
\centering
\begin{tabular}{ c >{\centering\arraybackslash}p{2cm} >{\centering\arraybackslash}p{2cm} >{\centering\arraybackslash}p{2cm}} 
\hline
\hline
& $Q_{{1A}}$-$Q_{{1B}}$ & $Q_{{2A}}$-$Q_{{2B}}$ &   $Q_{{3A}}$-$Q_{{3B}}$ \\
\hline 
$f_{|0_L\rangle}$ (GHz) & 4.311 & 3.826 & 4.136 \\
$T_{1,|0_L\rangle}$ ($\mu$s) & 59 & 63 & 41 \\
$f_{|1_L\rangle}$ (GHz) & 4.482 & 3.978 & 4.302 \\
$T_{1,|1_L\rangle}$ ($\mu$s) & 28 & 61 & 65 \\
$f_{L}$ (MHz) & 171.3 & 152.2 & 165.6 \\
$T_{1,L}$ (ms) & 0.80 & 1.15 & 1.01 \\
$T_{2,L}$ (ms) & 0.35 & 0.42 & 0.12 \\
\parbox[h]{2.6cm}{\setlength{\baselineskip}{10pt} readout fid.\\$F_{|0_L\rangle}$} & 97.5\% & 97.8\% & 97.8\% \\
$F_{|1_L\rangle}$ & 96.0\% & 98.1\% & 97.6\% \\
$F_{|00\rangle}$ & 99.6\% & 99.6\% & 99.3\%\\
\parbox[h]{2.6cm}{\setlength{\baselineskip}{10pt} logical 1Q \\ gate fid.}   & 99.992\% & 99.996\% & 99.992\% \\
\hline
\hline
\end{tabular}
\caption{Parameters of dual-rail qubits for experimental results in main text Fig.~4.}
\label{Table4}
\end{table}

\clearpage

\section{Ancilla qubit spectra}

\begin{figure}[!h]
    \centering
    \includegraphics[width=6.5 in]{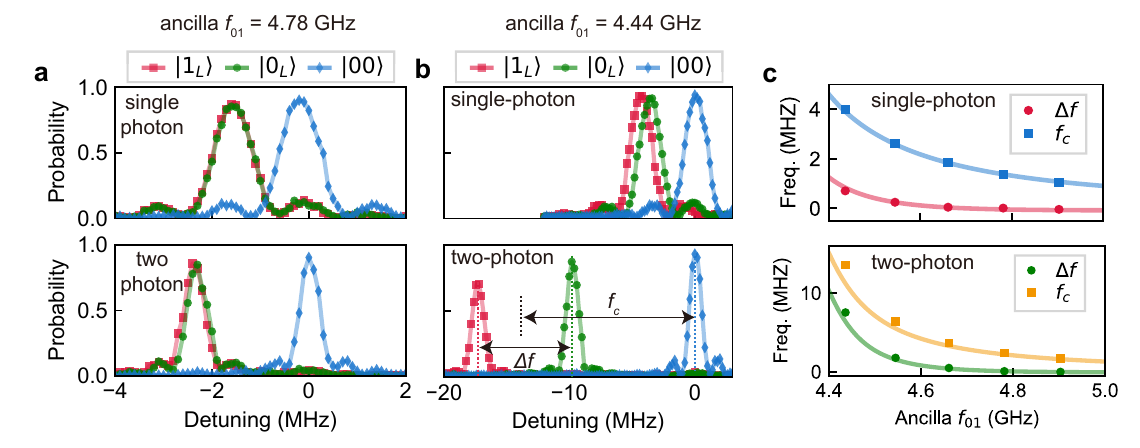}
    \caption{\label{FigSI:ancillaSpectra}
 \subfiglabel{a},
Single-photon and two-photon spectra of ancilla with ancilla frequency $f_{01} = 4.78$~GHz. The horizontal axes of the single-photon and two-photon spectra are the detuning with respect to $f_{01}$ and the detuning with respect to $f_{02}/2$, respectively.
 \subfiglabel{b},
Single-photon and two-photon spectra of ancilla with ancilla frequency $f_{01} = 4.44$~GHz.
 \subfiglabel{c},
Extracted frequency difference from different peaks in the ancilla single-photon and two-photon spectra as a function of ancilla frequency $f_{01}$. $\Delta f$ is the frequency difference between the peak $|0_L\rangle$ and peak $|1_L\rangle$. $f_c$ is the frequency difference between the peak $|00\rangle$ and the mean frequency of peak $|1_L\rangle$ and $|0_L\rangle$. 
Solid lines are simultaneous fit with the same fit parameters.
    }
\end{figure}

In Fig.~2 of the main text, we present the ancilla spectra for both single-photon and two-photon excitation schemes with dual-rail qubits in various states. Notably, the relative detuning between spectral peaks is strongly influenced by the ancilla frequency. Figures~\ref{FigSI:ancillaSpectra}~\textbf{a}-\textbf{b} illustrate the ancilla spectra at different ancilla frequencies, with the two data qubits of the dual-rail system tuned on resonance at 3.915~GHz. As the ancilla frequency approaches that of the data qubits, the dispersive interaction between them strengthens, resulting in larger detuning between spectral peaks. The frequency differences between these peaks, extracted from the single-photon and two-photon spectra, are shown in Figs.~\ref{FigSI:ancillaSpectra}~\textbf{c}-\textbf{d}, respectively.

To interpret these results, we compute the energy spectrum using the Hamiltonian of the dual-rail system. This system consists of three coupled qubits, with nonzero interactions between any two of them. Denoting the two data qubits as $A$ and $B$ and the ancilla as $a$, the Hamiltonian is given by:
\begin{equation}
\label{eq:SI_Hamiltonian}
\begin{aligned}
    H/\hbar =& \, \omega_{A} a^\dagger_{A} a_{A}+ \frac{\eta_{A}}{2}  a^\dagger_{A} a^\dagger_{A} a_{A} a_{A} + \omega_{B} a^\dagger_{B} a_{B} + \frac{\eta_{B}}{2}  a^\dagger_{B} a^\dagger_{B} a_{B} a_{B} \\
    & +\omega_{a} a^\dagger_{a} a_{a} + \frac{\eta_{a}}{2}  a^\dagger_{a} a^\dagger_{a} a_{a} a_{a} + g_{aA}(a^\dagger_{A} a_{a} + a^\dagger_{a} a_{A})  \\
    & + g_{aB}(a^\dagger_{B} a_{a} + a^\dagger_{a} a_{B}) + g_{AB}(a^\dagger_{A} a_{B} + a^\dagger_{B} a_{A}),
\end{aligned}
\end{equation}
where $a^\dagger_{i}$~($a_i$) represents the creation (annihilation) operator of qubit $i$, $\omega_{i}$ and $\eta_{i}$ is the frequency and anharmonicity of qubit $i$, $g_{ij}$ is the coupling strength between qubit $i$ and qubit $j$, and $\hbar$ is the reduced planck constant. For the data in Fig.~\ref{FigSI:ancillaSpectra}, the frequencies of the two data qubit satisfy $\omega_{1A}/2\pi = \omega_{1B}/2\pi = 3.915~\mathrm{GHz}$, with anharmonicities $\eta_{1A}/2\pi, \eta_{1B}/2\pi, \eta_{a}/2\pi \sim -0.2~\mathrm{GHz}$. Given that the capacitance between qubit pads is constant, the coupling strength scales as $g_{ij} \propto \sqrt{\omega_i \omega_j}$, and we treat the proportionality constants as fitting parameters to align with the experimental data. The dispersive shifts of the ancilla for different data qubit states are calculated by solving the eigenvalues of Hamiltonian~\eqref{eq:SI_Hamiltonian}. By fitting the experimental data, we obtain the coupling strength between the two data qubits to be $g_{AB} = 85$~MHz which reasonably agree with the measured value 76~MHz. This model provides a framework for selecting the ancilla frequency to achieve the desired spectral properties for erasure check using the ancilla.

\section{Erasure check fidelity}

\begin{figure}[!b]
    \centering
    \includegraphics[width=0.9\textwidth]{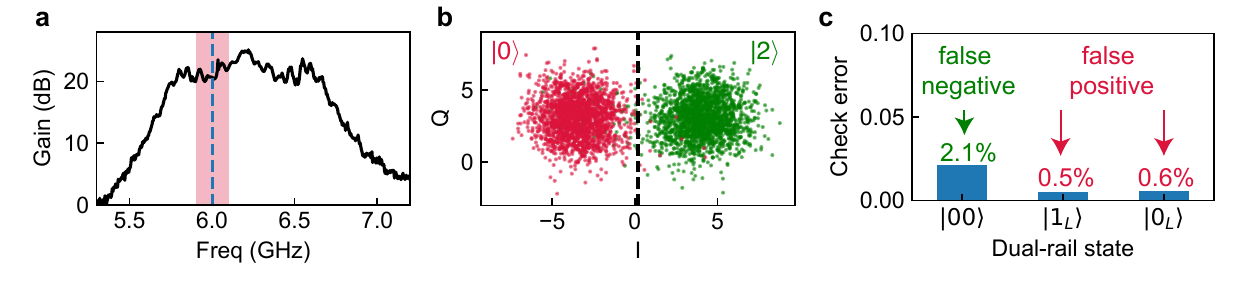}
    \caption{\label{FigSI:erasureCheck}
 \subfiglabel{a},
Gain profile of the IMPA used for ancilla readout. The colored region marks the readout frequency range for different ancilla. The dashed line marks the readout frequency of the ancilla used for the data in (b) and (c).
 \subfiglabel{b},
Distribution of the I-Q cloud of ancilla measurement for the ground state $|0\rangle$ and two-photon state $|2\rangle$. The center of the two cloud is aligned with the I axis. The dashed line marks the I value to discriminate the two states.
  \subfiglabel{c},
Error of ancilla check with dual-rail qubit prepared in different states. With the dual-rail qubit prepared in $|00\rangle$ state, which represents the case of erasure error and the ancilla state should read as $|2\rangle$ during the check, the corresponding check error (false negative error) is the probability the ancilla reads state $|0\rangle$. With the dual-rail qubit prepared in $|0\rangle_L$ or $|1\rangle_L$ state, which represents the case of no erasure error and the ancilla state should be $|0\rangle$ during the check, the corresponding check error (false positive error) is the probability the ancilla reads state $|2\rangle$.
    }
\end{figure}

To achieve reliable erasure checking with the ancilla, high readout fidelity is essential. We employ a quantum-limited IMPA as the preamplifier for the readout line of the ancillas. Figure~\ref{FigSI:erasureCheck}~\textbf{a} presents the measured gain profile of the IMPA. The gain is approximately 20 dB over a frequency range of ~1 GHz, which covers all four ancilla qubits simultaneously.

To benchmark the readout quality, we perform standard qubit measurements on the ancilla and also evaluate the check error on the dual-rail qubit using the ancilla. Figure~\ref{FigSI:erasureCheck}~\textbf{b} shows the measured I-Q distribution of the readout signal with the ancilla in the $|0\rangle$ and $|2\rangle$ states respectively. Here, the state $|2\rangle$ is used as the excited state, as a two-photon excitation scheme is employed for the ancilla check in our experiments. The histogram of the I-Q distribution reveals a signal-to-noise ratio (SNR) of 3.8. Here, the SNR is defined as $\mathrm{SNR}^2 = (c_2 - c_0)^2 / (\sigma_2^2 + \sigma_0^2)$, where $c_i$ and $\sigma_i$ represent the center and standard deviation of the Gaussian distributions of the I-Q cloud for state $|i\rangle$. The center $c_i$ corresponds to the average I value, as the I-Q clouds for the two states are rotated to align with the I-axis. The SNR of 3.8 corresponds to a separation error of 0.3\%, which sets an upper bound for the readout fidelity at 99.7\%. The measured average readout fidelity is 97.4\%, primarily limited by state preparation error and $T_1$ relaxation during the readout pulse.

Figure~\ref{FigSI:erasureCheck}~\textbf{c} shows the measured check error. These results are obtained by first preparing the dual-rail qubit in the $|00\rangle$, $|0_L\rangle$, and $|1_L\rangle$ states, followed by an erasure check using the ancilla. If the dual-rail qubit is in the logical space ($|0_L\rangle$ or $|1_L\rangle$), the ancilla check should read as $|0\rangle$, and the corresponding readout error is classified as a false positive error, representing the case where no erasure error occurred, but an erasure was falsely identified. Conversely, if an erasure error occurs thus the dual-rail qubit is in the $|00\rangle$ state, the ancilla check should read as $|2\rangle$, and the corresponding readout error is a false negative error, representing the case where an erasure error occurred, but no error was correctly identified. To more accurately characterize the check errors, an additional end measurement is applied to the dual-rail qubit. By performing a post selection such that only the shots where the final measurement of the dual-rail qubit matches the prepared state are retained, this ensures that $T_1$ relaxation of the dual-rail qubit does not affect the measured check error. The false negative error is measured to be 2.1\%, while the false positive error is approximately 0.5\%. The higher false negative error is attributed to the ancilla being in the excited state $|2\rangle$ during the readout pulse, making it more susceptible to $T_1$ relaxation effects.


\section{Dual-rail qubit coherence with ancilla check}

The primary error of the dual rail qubits are the erasure error due to $T_1$ relaxation of the physical qubits which makes the logical states $|0_L\rangle$ and $|1_L\rangle$ decay to the ground state $|00\rangle$. Ancilla check and also the end measurement of the dual-rail qubit could all detect this error and mitigate it by post selection. There is also another mechanism that could induce decoherence, i.e. the slow heating of the qubits which could bring the population from $|00\rangle$ back to the logical space. The dual-rail qubit could decay from the logical space to $|00\rangle$ and then reheat back to the logical space without being detected which effectively induce dephasing of the logical qubit. 

\begin{figure}[!h]
    \centering
    \includegraphics[width=6.5 in]{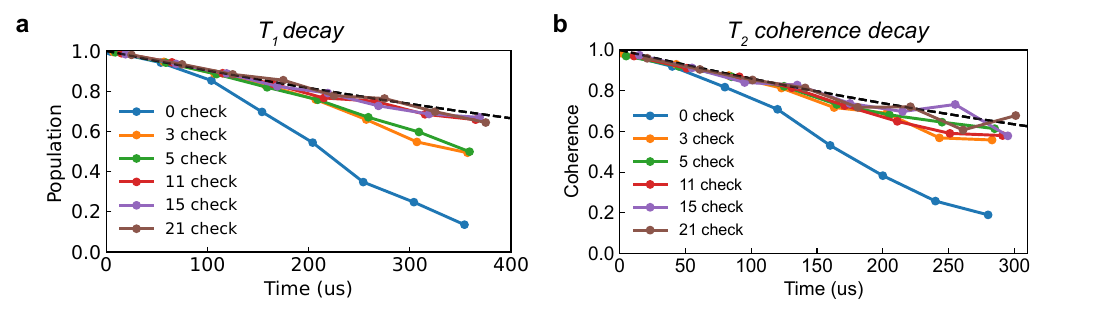}
    \caption{\label{FigSI:T1T2_vs_check}
    \subfiglabel{a},
    $T_1$ decay in the logical subspace as a function of the check number. Dashed line is an exponential decay using $T_1 = 0.98$~ms, which is the same as the value in Fig.~2\textbf{c} in the main text.
    \subfiglabel{b},
    $T_2$ coherence decay in the logical subspace as a function of the check number, measured with CPMG 256 sequence. Coherence of 1 represents the maximum value of the off-diagonal elements of the density matrix for an initial state of $(|0\rangle_L+|1\rangle_L)/\sqrt{2}$. Dashed line is an exponential decay using $T_2 = 0.66$~ms, which is the same as the value in Fig.~2\textbf{d} in the main text.
    }
\end{figure}

\begin{figure}[h]
    \centering
    \includegraphics[width=3 in]{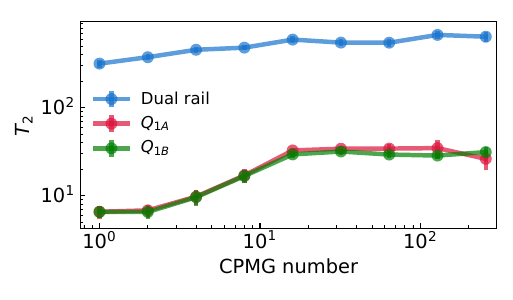}
    \caption{\label{FigSI:T2_vs_CPMG}
    $T_2$ as a function of the pulse number in CPMG sequence for dual-rail qubit and the two physical qubit respectively.
}
\end{figure}

This decoherence mechanism could be effectively mitigated by frequently performing ancilla check during the experimental sequence. By performing multiple ancilla check and only keeping the events where all the checks show no erasure error, the possibility of the reheating event becomes small. Figure~\ref{FigSI:T1T2_vs_check} shows how the coherence decays during $T_1$ and $T_2$ decay process. By repeatdly applying ancilla checks, the coherence can last much longer than the case with zero ancilla check. Up to a threshold, increasing check number does not have more benefit. Utilizing this, by performing 21 checks, one order of magnitude enhancement of $T_1$ is achieved in the dual-rail qubit compared to its physical counterpart. Additionally, by employing various CPMG sequences, we achieve an enhancement in $T_2$ by one to two orders of magnitude, as illustrated in Fig.~\ref{FigSI:T2_vs_CPMG}.

\clearpage

\section{Calibration of logical single qubit gates}

\begin{figure}[!b]
    \centering
    \includegraphics[width=5 in]{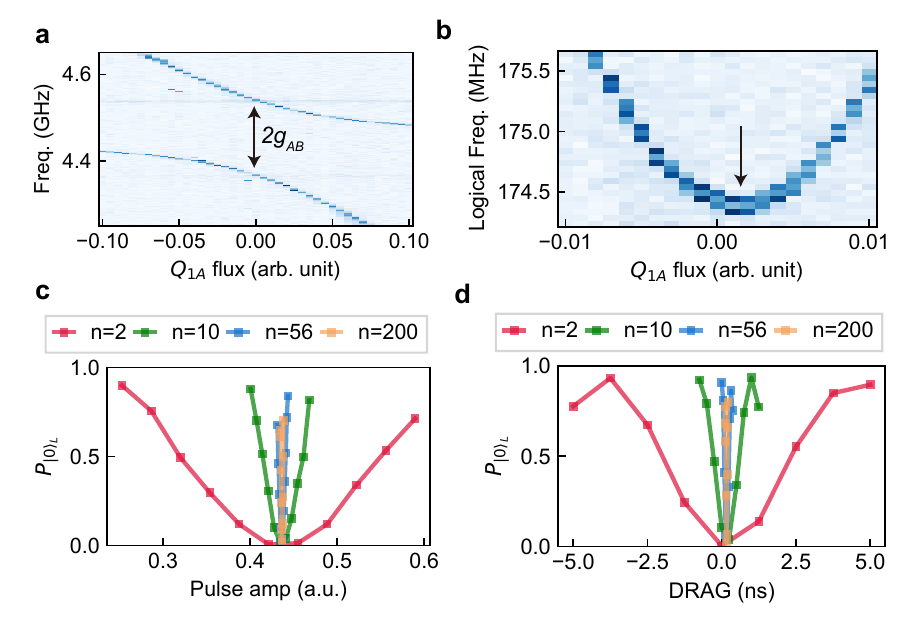}
    \caption{\label{FigSI:singleGateCali}
 \subfiglabel{a},
Spectroscopy of the two physical qubits in a dual-rail system as a function of the flux of physical qubit $Q_{1A}$.
 \subfiglabel{b},
Logical spectroscopy of a single dual-rail qubit as a function of the flux of physical qubit $Q_{1A}$. The arrow points to the qubit flux where the logical frequency is the minimum.
  \subfiglabel{c},
Calibration of the pulse amplitude of a logical $\pi$ pulse, where $n$ is the number of repeated gates in the calibration sequence.
  \subfiglabel{d},
Calibration of the DRAG parameter of a logical $\pi$ pulse.
    }
\end{figure}

The calibration of logical single qubit gates involves three steps:

1. \textbf{Tune the two data qubits to resonance.} When the data qubits are brought into resonance, an avoided crossing becomes clearly visible in qubit spectroscopy, as shown in Fig.\ref{FigSI:singleGateCali}\textbf{a}. For this data, the corresponding splitting, $2g_{AB}$, is approximately 175~MHz. Notably, this splitting varies with the resonant frequency of the data qubits, following the relation $g_{AB} \propto \sqrt{\omega_A \omega_B}$. The qubit spectroscopy is obtained by initializing the qubit in $|0\rangle$, then applying a long drive pulse ($>40\mu s$) at various frequencies and qubit biases, and then measuring the qubit population.

2. \textbf{Fine tuning of the logical frequency.} The logical qubit spectroscopy at various qubit biases is shown in Fig.\ref{FigSI:singleGateCali}\textbf{b}. The data is obtained by initializing the logical qubit in either $|0_L\rangle$ or $|1_L\rangle$, applying a long drive pulse ($>40~\mu s$) at various frequencies near 175~MHz and with different qubit biases, and then measuring the population of the logical states. The minimum logical frequency and its corresponding bias point are identified, providing a more robust operating point against flux noise and improving the logical qubit coherence.

\begin{figure}[!t]
    \centering
    \includegraphics[width=3 in]{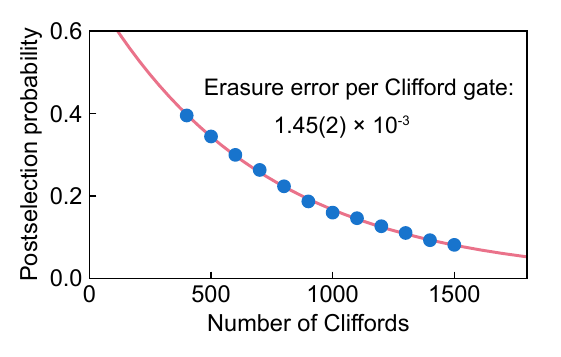}
    \caption{\label{FigSI:singleGateProb}
The postselection probability as a function of the Clifford gate numbers for during the logical reference randomized benchmarking measurements. The solid line is an exponential fit, which is used to extract the erasure error per Clifford gate. Each Clifford gate consists of 2.25 $\pi/2$ pulses on average thus the erasure error per $\pi/2$ pulse is $6.4(1)\times10^{-4}$. This data serves as the supplementary data for Fig.~2\textbf{e} in the main text.
    }
\end{figure}

3. \textbf{Calibrating the pulse amplitude and DRAG parameter.} The amplitude and the derivative reduction by adiabatic gate (DRAG) parameter are calibrated using repeated pulses. Figures~\ref{FigSI:singleGateCali}~\textbf{c-d} present calibration data for a logical $\pi$ pulse. The procedure involves initializing the logical qubit in $|0_L\rangle$, applying $n$ repeated $\pi$ pulses with varying amplitudes and DRAG parameters, and determining the optimal parameter set by minimizing $P_{|0_L\rangle}$. The calibration of a logical $\pi/2$ pulse follows a similar approach. Increasing the number of pulses gradually improves calibration accuracy. In our calibration, optimal parameters are determined at $n \sim 200$. After the calibration, we use randomized benchmarking in the logical space to determine the logical single-qubit gate error and the erasure error, with typical gate error data shown in main text Fig.~2\textbf{e} and typical erasure error data shown in Fig.~\ref{FigSI:singleGateProb}.

The DRAG method refines pulse control by adjusting the two quadratures so that the envelope of the second quadrature is the time derivative of that of the first. We define the drag parameter of the logical pulse in the following way. Assuming an uncorrected pulse $f(t)=s(t)\cos(\omega t+\phi)$ is applied to the dual-rail qubit, the corresponding DRAG pulse is $f'(t) = s(t)\cos(\omega t + \phi) + \beta \dot{s}(t)\sin(\omega t + \phi)$, where $s(t)$ is the envolope of the pulse, $\omega$ is the logical frequency of the dual-rail qubit, $\phi$ is the phase of the pulse, and $\beta$ is the DRAG parameter.

\clearpage

\section{Calibration of logical two qubit gates} \label{sec:cali_twoQ}
\subsection{Energy level shift while tuning the coupler}

\begin{figure}[!b]
    \centering
    \includegraphics[width=5 in]{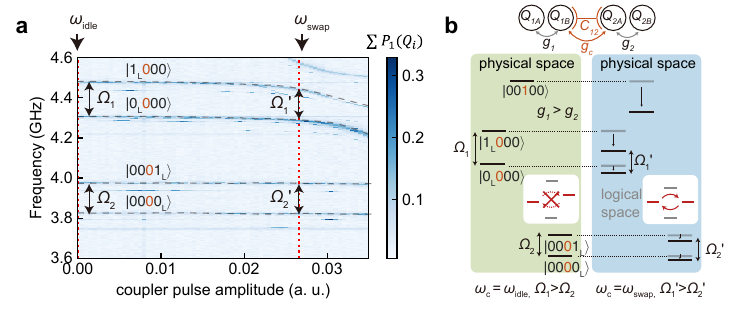}
    \caption{\label{FigSI:spec_vs_cpa}
    \textbf{a,} Qubit spectroscopy as a function of the coupler pulse amplitude. The plotted figure is the combination of the spectra of all the physical qubits in the two dual-rail qubits. \textbf{b,} Schematics of the energy level shift with the coupler at the idle frequency $\omega_{\mathrm{idle}}$ and the swap frequency $\omega_{\mathrm{swap}}$. In our notation, $|00100\rangle$ represents the composite state of two dual-rail qubits and the coupler: the first (second) segment—denoted as $|00\rangle$ (or $|0_L\rangle$, $|1_L\rangle$)—indicates the state of the first (second) dual-rail qubit, while the middle $|1\rangle$ corresponds to the coupler's state.
}
\end{figure}

The logical $\sqrt{i\mathrm{SWAP}}$ gate is implemented using a single flux tuning pulse applied to the coupler. The coupler’s idling frequency, $\omega_{\mathrm{idle}}$, is chosen such that the $ZZ$ interaction between the two logical qubits is eliminated, preventing unwanted accumulated $ZZ$ phase. At this idling point, the logical qubit frequencies, $\Omega_1$ and $\Omega_2$, are distinct, with $\Omega_1 > \Omega_2$. Additionally, $\omega_{\mathrm{idle}}$ is set higher than the frequencies of all four physical qubits.

During gate operation, the coupler frequency is tuned down to $\omega_{\mathrm{swap}}$, which is close to the highest frequency among the four physical qubits. Due to state hybridization, this tuning also lowers the frequencies of all four physical qubits, with a stronger effect on the first dual-rail qubit, which has the logical frequency $\Omega_1$. With the coupler frequency tuned to $\omega_{\mathrm{swap}}$, the two dual-rail qubit frequencies become equal, i.e., $\Omega_1' = \Omega_2'$. This behavior is clearly illustrated in the qubit spectroscopy data, which maps the physical qubit frequencies as the coupler is tuned between $\omega_{\mathrm{idle}}$ and $\omega_{\mathrm{swap}}$, as shown in Fig.\ref{FigSI:spec_vs_cpa}~\textbf{a-b}.

\begin{figure}[!t]
    \centering
    \includegraphics[width=6 in]{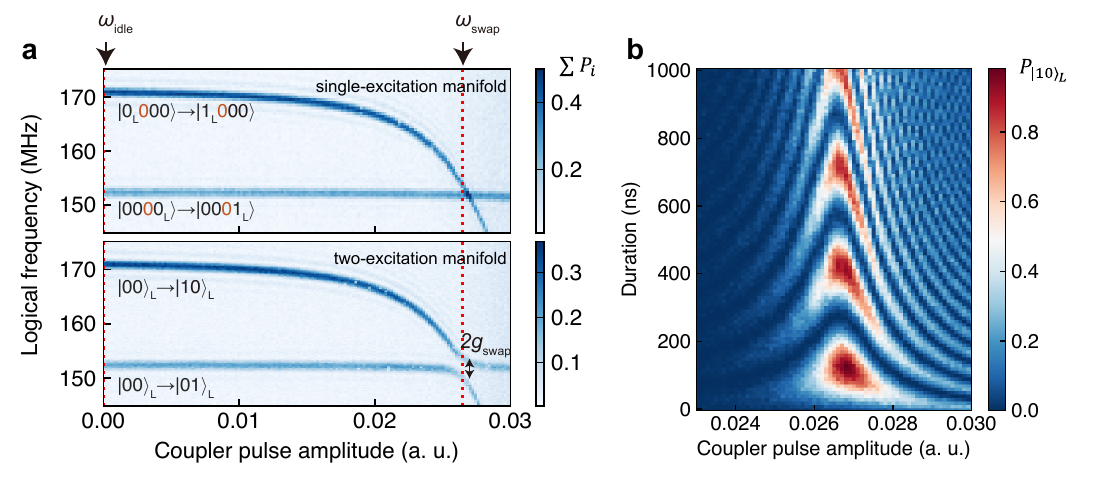}
    \caption{\label{FigSI:logical_spec_vs_cpa}
    \textbf{a,} Logical qubit spectroscopy as a function of the coupler pulse amplitude. The plotted figure is the combination of the spectra of the two transitions labeled in each panel. \textbf{b,} Logical Rabi oscillation between $|01\rangle_L$ and $|10\rangle_L$ as function of the coupler pulse amplitude.
}
\end{figure}

To verify the presence of a logical $XX$ interaction at $\omega_{\mathrm{swap}}$, we perform logical qubit spectroscopy on the two dual-rail qubits, as shown in Fig.\ref{FigSI:logical_spec_vs_cpa}~\textbf{a}. The data confirm that $\Omega_1 > \Omega_2$ with coupler at $\omega_{\mathrm{idle}}$ and that they become degenerate ($\Omega_1' = \Omega_2'$) with coupler at $\omega_{\mathrm{swap}}$. Furthermore, in the two-excitation manifold, where both dual-rail qubits reside in the logical space, we observe an avoided crossing with a gap of several MHz, confirming the presence of a nonzero logical $XX$ interaction with coupler at $\omega_{\mathrm{swap}}$. Notably, this avoided crossing is absent in the single-excitation manifold, where only one dual-rail qubit is in the logical space while the other remains in the $|00\rangle$ state, indicating that the logical two-qubit interaction is effective only when both qubits are within the logical space.

With coupler frequency near $\omega_{\mathrm{swap}}$, coherent population exchange between $|01\rangle_L$ and $|10\rangle_L$ is clearly observed, along with a pronounced chevron pattern as the coupler flux is varied, as shown in Fig.\ref{FigSI:logical_spec_vs_cpa}~\textbf{b}. By selecting the coupler flux that maximizes oscillation visibility, we determine the parameters used in Fig.~3\textbf{b} of the main text. The pulse duration required to implement a $\sqrt{i\mathrm{SWAP}}$ gate is identified at the point where the populations of $|01\rangle_L$ and $|10\rangle_L$ become equal.

\subsection{Accumulated phase calibration and correction}

Since the logical $\sqrt{i\mathrm{SWAP}}$ gate is implemented by tuning the coupler frequency, it inevitably introduces accumulated phases on the two logical qubits. Accounting for these accumulated phases, the matrix representation of the logical $\sqrt{i\mathrm{SWAP}}$ gate is given by:
\begin{equation}
U = \begin{pmatrix}
1 & 0 & 0 & 0 \\
0 & e^{i\phi_2}/\sqrt{2} & -ie^{i(\phi_1-\delta)}/\sqrt{2} & 0 \\
0 & -ie^{i(\phi_2+\delta)}/\sqrt{2} & e^{i\phi_1}/\sqrt{2} & 0 \\
0 & 0 & 0 & e^{i(\phi_1+\phi_2+\phi_{zz})}
\end{pmatrix},
\end{equation}
where $\phi_1$ and $\phi_2$ are the accumulated single qubit phases of the two logical qubits, $\delta$ is the accumulated relative phase due to their different idling frequencies, and $\phi_{zz}$ is the accumulated controlled phase caused by the nonzero $ZZ$ coupling during the logical $\sqrt{i\mathrm{SWAP}}$ gate.

\begin{figure}[!b]
    \centering
    \includegraphics[width=5 in]{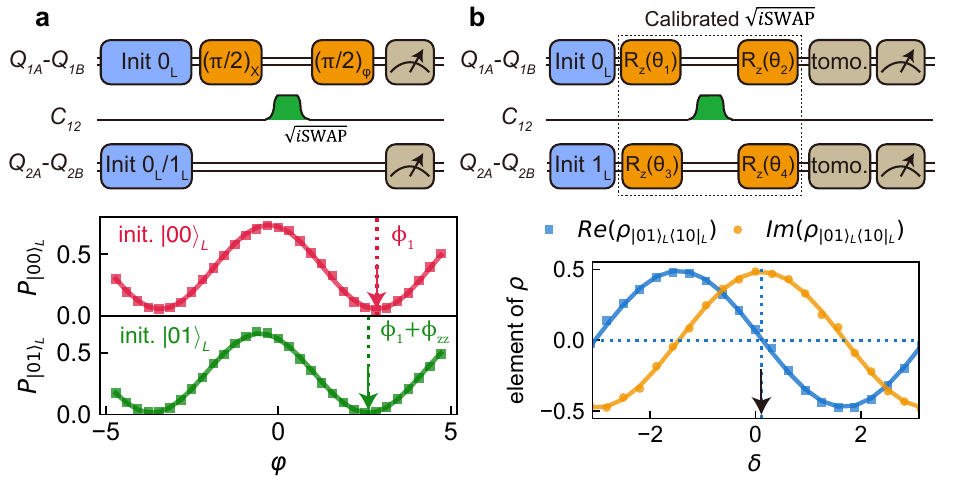}
    \caption{\label{FigSI:twoQubitCali}
 \subfiglabel{a},
Calibration of accumulated single qubit phase during $\sqrt{i\mathrm{SWAP}}$ gate. The top pannel is the sequence and the bottom panel is the measured data. The measured data are the population of $|00\rangle_L$ ($|01\rangle_L$) as a function of the rotation angle of the second $\pi/2$ pulse with the logical qubit first initialized to $|00\rangle_L$ ($|01\rangle_L$). Solid lines are sinusoid fit. The presented data and pulses sequence is for calibration of the accumulated phase in the first dual-rail qubit. The arrows mark the determined calibrated phases. 
 \subfiglabel{b},
Calibration of the relative phase $\delta$ during  $\sqrt{i\mathrm{SWAP}}$ gate. The top pannel is the sequence and the bottom panel is the measured data. The measured data are the off-diagonal term of the density matrix, $\rho_{|01\rangle_L\langle10|_L}$, as a function of the relative phase $\delta$ used in the four calibration pulses. Solid lines are sinusoid fit. The arrow marks the determined calibrated phase.
    }
\end{figure}

\begin{figure}[!t]
    \centering
    \includegraphics[width=3 in]{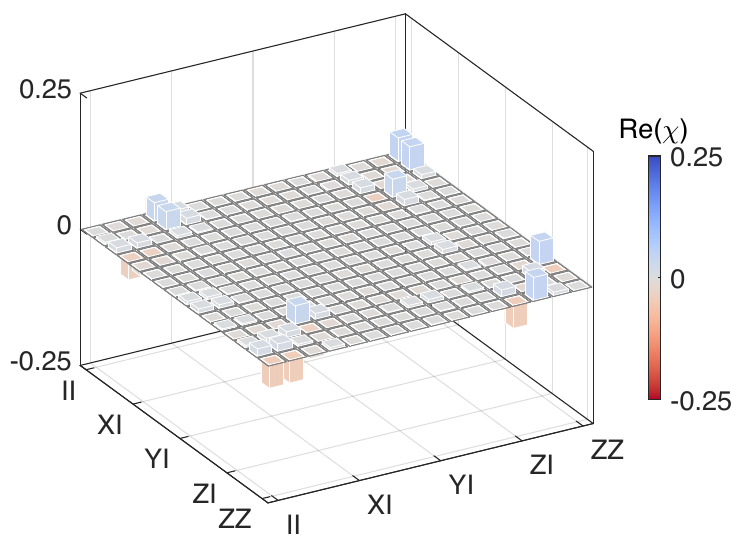}
    \caption{\label{FigSI:CNOT_imag}
The imaginary part of the process matrix $\chi$ for the logical CNOT gate, serving as supplementary data to Fig.~4\textbf{b} in the main text.
    }
\end{figure}

The single-qubit phases and the relative phase can be corrected using four logical $R_z$ gates, as illustrated in Fig.~\ref{FigSI:twoQubitCali}~\textbf{b}. In our experiment, the logical $R_z$ gate is constructed by two $\pi$ pulses: $R_z(\theta) = R_{-\theta/2}(\pi)R_x(\pi)$. The rotation angles for the four $R_z$ gates are determined as follows:
\begin{equation}
\label{eq:single_gate_cali}
\begin{aligned}
\theta_1 &= 3\phi_1/4 - \phi_2/4 - \delta/2, \\
\theta_2 &= \phi_1/4 + \phi_2/4 + \delta/2 + \phi_{zz}/2, \\
\theta_3 &= -\phi_1/4 + 3 \phi_2/4 + \delta/2, \\
\theta_4 &= \phi_1/4 + \phi_2/4 - \delta/2 + \phi_{zz}/2,
\end{aligned}    
\end{equation}
where $R_z(\theta_1)$ and $R_z(\theta_3)$ are applied before the coupler flux pulse to the first and second dual-rail qubits respectively, and $R_z(\theta_2)$ and $R_z(\theta_4)$ are applied after the coupler flux pulse to the first and second dual-rail qubits respectively. With this correction, the calibrated logical $\sqrt{i\mathrm{SWAP}}$ gate changes to:
\begin{equation}
U = \begin{pmatrix}
1 & 0 & 0 & 0 \\
0 & 1/\sqrt{2} & -i/\sqrt{2} & 0 \\
0 & -i/\sqrt{2} & 1/\sqrt{2} & 0 \\
0 & 0 & 0 & e^{-i\phi_{zz}}
\end{pmatrix},
\end{equation}
where the controlled phase $\phi_{zz}$ remains in the operation. To eliminate this, we construct the logical CNOT gate by using two logical $\sqrt{i\mathrm{SWAP}}$ gates interleaved by a logical $X$ pulse and also with a few other single qubit gates, as elucidated in the main text and the corresponding process matrix of the CNOT gate is shown in main text Fig.~4\textbf{b} (real part) and Fig.~\ref{FigSI:CNOT_imag} (imaginary part).

The single-qubit phases $\phi_{1,2}$ are calibrated using a Ramsey sequence. By initializing both logical qubits in $|00\rangle_L$ and applying a Ramsey sequence to the first dual-rail qubit with the coupler flux pulse inserted between two $\pi/2$ pulses, the qubit population oscillates as a function of the phase $\phi$ of the second $\pi/2$ pulse. The phase $\phi_1$ is extracted by fitting the data to determine the phase where $P_{|00\rangle_L}$ is minimized. Similarly, by initializing the logical qubits in $|01\rangle_L$ and repeating the procedure, the extracted phase corresponds to $\phi_1 + \phi_{zz}$, confirming the presence of a nonzero accumulated controlled phase $\phi_{zz}$. The phase $\phi_2$ is obtained in the same manner by applying the Ramsey sequence to the second dual-rail qubit.

Once the single-qubit phases $\phi_{1,2}$ and the controlled phase $\phi_{zz}$ are calibrated, the relative phase $\delta$ is determined using quantum state tomography (QST) of the logical Bell state generated by the $\sqrt{i\mathrm{SWAP}}$ gate, as shown in Fig.\ref{FigSI:twoQubitCali}~\textbf{b}. Starting with the logical qubits in $|01\rangle_L$, the calibrated $\sqrt{i\mathrm{SWAP}}$ gate (corrected as per Eq.~\eqref{eq:single_gate_cali}) is applied, followed by QST on the resulting Bell state $\rho$ for different calibration values of $\delta$. The optimal $\delta$ is chosen such that the off-diagonal term $\rho_{|01\rangle_L\langle10|_L}$ is a positive imaginary number, aligning with the expected Bell state generated by an ideal $\sqrt{i\mathrm{SWAP}}$ gate.

\clearpage
\section{Error analysis of logical gates}

For single-qubit gates, the coherence-limited gate error~\cite{o2015qubit} is given by $\varepsilon = \tau/3T_1 + \tau/3T_\phi$, where $\tau$ is the pulse duration, $T_1$ is the relaxation time, and $T_\phi$ is the pure dephasing time due to white noise. Based on the randomized benchmarking (RB) data in Fig.~2\textbf{d} of the main text, we have $\tau = 25$~ns, $T_1 = 980$~$\mu$s, and $T_\phi = 1/(1/T_1 - 1/T_{2,CPMG}) = 995$~$\mu$s. This yields a coherence-limited gate error of $1.7\times10^{-5}$, which is close to the measured value of $2.6\times10^{-5}$. 

The error of the logical CNOT gate is approximately 3.8\%, significantly larger than that of single-qubit gates. As discussed in Sec.\ref{sec:cali_twoQ}, during the $\sqrt{i\mathrm{SWAP}}$ gate, the coupler frequency is tuned close to the physical qubits $Q_{1A}$ and $Q_{1B}$, leading to strong hybridization and increased decoherence. The decoherence times of the logical qubits at the idle frequency $\omega_{\mathrm{idle}}$ and the interaction frequency $\omega_{\mathrm{swap}}$ are summarized in Table~\ref{Table:coupler_dephasing}. The $T_{1,L}$ and $T_{2,L}$ of the first dual-rail qubit decrease from several hundreds of $\mu$s to just a few $\mu$s. This effect is less pronounced for the second dual-rail qubit, as its physical qubits operate at lower frequencies.

The logical CNOT gate comprises two $\sqrt{i\mathrm{SWAP}}$ operations, each implemented via a coupler flux pulse featuring a ~50~ns plateau flanked by two 50~ns edge segments (see Fig.~3a of the main text for the shape of the pulse), which adiabatically tune the coupler frequency from $\omega_{\mathrm{idle}}$ to $\omega_{\mathrm{swap}}$. The coherence limited gate error of the two qubit gate is mainly determined by $T_1$ relaxation and white noise dephasing $T_{\phi}$: $\varepsilon = \sum_i 2/5(1/T_{1,Q_i}+1/T_{\phi,Q_i}) \tau$~\cite{sete2024error}, where $T_{1,Q_i}$ and $T_{\phi,Q_i}$ are the relaxation time and dephasing time of the $i$th dual-rail qubit while the coupler is at $\omega_{\mathrm{swap}}$, and $\tau$ is the duration of the pulse. Given the challenges in accurately determining decoherence times during the pulse edges, we estimate the gate error bounds by considering only the plateau's decoherence effects for the lower bound ($\tau=50$~ns) and assuming uniform decoherence times during both plateau and edges for the upper bound ($\tau=150$~ns). This approach yields an estimated error of 0.8\%–2.5\% for a single $\sqrt{i\mathrm{SWAP}}$ gate, culminating in a logical CNOT error estimate of 1.7\%–5\%, aligning reasonably with the observed gate error of 3.8\%. The predominant contributor to this substantial gate error is the coupler-induced decoherence of the dual-rail qubits. Using this estimation method, maintaining the dual-rail qubits' coherence at idling values could elevate gate fidelity beyond 99.9\%. Future efforts will focus on optimizing coupler designs and refining pulse sequences to achieve this objective.


\begin{table}[h!]
\centering
\begin{tabular}{ >{\centering\arraybackslash}p{5cm} >{\centering\arraybackslash}p{3cm} >{\centering\arraybackslash}p{3cm} } 
\hline
\hline
& $T_{1,L}$ (ms)  & $T_{2,L}$ (ms) \\
\hline 
$Q_{{1A}}$-$Q_{{1B}}$ @ $\omega_c = \omega_{\mathrm{idle}}$   & 0.59 & 0.24 \\
$Q_{{1A}}$-$Q_{{1B}}$ @ $\omega_c = \omega_{\mathrm{swap}}$ & 0.003 & 0.004 \\
$Q_{{2A}}$-$Q_{{2B}}$ @ $\omega_c = \omega_{\mathrm{idle}}$   & 1.29 & 0.62 \\
$Q_{{2A}}$-$Q_{{2B}}$ @ $\omega_c = \omega_{\mathrm{swap}}$ & 0.35 & 0.17 \\
\hline
\end{tabular}
\caption{Comparison of logical coherence time of two coupled dual-rail qubits while the coupler is at $\omega_{\mathrm{idle}}$ and $\omega_{\mathrm{swap}}$.}
\label{Table:coupler_dephasing}
\end{table}


\bibliographystyle{naturemag}
\bibliography{ref_main}